\documentclass[a4paper,11pt]{article}
\usepackage{jheppub}
\usepackage{dsfont}
\usepackage{fontenc}
\usepackage{slashed}
\usepackage{amsmath}
\usepackage{amssymb}
\usepackage{bm}

\def\twotwo{2\leftrightarrow 2}
\def\sss{\scriptscriptstyle}

\def \cf {C_{\sss F}}
\def \nc {N_c}
\def \ca {C_{\sss A}}
\def \nf {N_f}

\def \dr {d_{\sss R}}

\def \bk {\mathbf{k}}

\def \bp {\mathbf{p}}

\newcommand{\Tint}[1]{{\hbox{$\sum$}\!\!\!\!\!\!\!\int\,}_{\!\!\!\!\raise-0.9ex\hbox{$\scriptstyle{#1}$}}}

\def \cc{{\mathcal C}}

\def \Bb{{\mathcal B}}

\def\siml{{\ \lower-1.2pt\vbox{\hbox{\rlap{$<$}\lower6pt\vbox{\hbox{$\sim$}}}}\ }}
\def\simg{{\ \lower-1.2pt\vbox{\hbox{\rlap{$>$}\lower6pt\vbox{\hbox{$\sim$}}}}\ }}

\def \crr {C_{\sss R}}

\def \bk {\mathbf{k}}

\def \bp {\mathbf{p}}

\def \m2   {\mu^{2 \epsilon}}

\newcommand{\order}[1]{\mathcal{O}\left(#1\right)}

\def\siml{{\ \lower-1.2pt\vbox{\hbox{\rlap{$<$}\lower6pt\vbox{\hbox{$\sim$}}}}\ }}
\def\simg{{\ \lower-1.2pt\vbox{\hbox{\rlap{$>$}\lower6pt\vbox{\hbox{$\sim$}}}}\ }}

\def\nn {\nonumber}

\def\pp {p_\perp}

\def\qp {q_\perp}
\def\bpp {\mathbf{p}_\perp}
\def\bqp {\mathbf{q}_\perp}

\def\justm {m_\infty}
\def\mm {m_\infty^2}
\def\md {m_{\sss D}}
\def\Ccoll{C_\mathrm{coll}}

\def\2to2{{2\leftrightarrow 2}}

\def\hard{{\mathrm{hard}}}
\def\coll{{\mathrm{coll}}}

\def\alphas{\alpha_{\mathrm s}}
\def\Eq#1{Eq.~\eqref{#1}}
\def\Fig#1{Fig.~\ref{#1}}

\def\k{{\bf{k}}}
\def\f{{\bf{f}}}
\def\q{{\bf{q}}}
\def\b{{\bf{b}}}

\def\ell{{\bf{l}}}

\def\OO{{\mathcal{O}}}
\def\gsim{\mbox{~{\raisebox{0.4ex}{$>$}}\hspace{-1.1em}
    {\raisebox{-0.6ex}{$\sim$}}~}}

\def\nfd{n_{\!\sss F}}
\def\nbe{n_{\!\sss B}}

\title{Low Mass Thermal Dilepton Production at NLO
  in a Weakly Coupled Quark-Gluon Plasma}
\author[1,2]{Jacopo Ghiglieri,}
\author[1]{Guy D. Moore}

\affiliation[1]{McGill University, Department of Physics, 
  3600 rue University, Montreal QC H3A 2T8, Canada}
\affiliation[2]{Institute for Theoretical Physics, Albert Einstein
  Center, University of Bern, Sidlerstrasse 5, 3012 Bern, Switzerland} 
\emailAdd{jacopo.ghiglieri@itp.unibe.ch}
\emailAdd{guymoore@physics.mcgill.ca}

\abstract{
We present a computation, within weakly-coupled thermal QCD,
of the production rate of low invariant mass ($M^2 \sim g^2 T^2$)
dileptons, at next-to-leading order (NLO) in the coupling
(which is $\OO(g^3 e^2 T^2)$).  This involves extending the NLO
calculation of the photon rate which we recently presented to the case
of small nonzero photon invariant mass.
Numerical results are discussed and tabulated forms and code are
provided for inclusion in hydrodynamical models.
We find that NLO corrections can increase the dilepton
rate by up to 30-40\% relative to leading order.
We find that the electromagnetic response of the plasma for real photons
and for small invariant mass but high energy dilepton pairs
(\textsl{e.g.}, $M^2 < (300\:\mathrm{MeV})^2$ but $p_T > 1 \:
\mathrm{GeV}$) are close enough that dilepton pair measurements really
can serve as \textsl{ersatz} photon measurements.
We also present a matching \textsl{a la} Ghisoiu and Laine between our
results and results at larger invariant masses.
}

\keywords{Dileptons, Hard Probes, Quark-Gluon Plasma, High order
calculations}

\begin{document}
\maketitle

\flushbottom

\section{Introduction}

Most of the particles originating from an ultra-relativistic heavy ion
collision are believed
to undergo significant ``rescattering'' (parton energy loss, scattering,
fragmentation, hadronization, hadron-hadron scattering) before they
emerge and fly to the detector.  This degrades the information they
carry about the early stages of the collision, which must be
reconstructed by trying to model the scattering processes, for instance
with hydrodynamics.
Hard probes, on the other hand, are anything sufficiently high-energy or
weakly-coupled that they can penetrate the heavy ion environment with
little re-interaction.  They therefore carry more direct and
unprocessed information about early conditions.  Electromagnetic probes
are a good example, because the electromagnetic coupling strength
$\alpha = 1/137$ is small enough that re-interactions are rare.

Photons are one hard probe candidate.  With this in mind we recently
performed an improved determination of the rate of photon production
from a (weakly-coupled) quark-gluon plasma \cite{Ghiglieri:2013gia}.  But
photons come with some experimental challenges.  The highest energy
photons are expected to come primarily from collisions between the
initial state partons \cite{highEphotons}, and therefore carry little
information about the QGP evolution.  At intermediate energies where the
Quark-Gluon Plasma is expected to contribute (say, 1 to 4 GeV),
a photon excess has been observed \cite{highEphotons}.  These
measurements are challenging because of the large background of decay
photons which must be subtracted.  For
this reason, experimentalists have also focused on small-mass
dileptons, which we can think of as massive off-shell photons.
Provided the mass-squared of the pair is above the pion mass, the
(Dalitz) pion decay background is absent and the foreground rates are
under much better control.  For this reason, $e^+ e^-$ pairs
with invariant masses somewhat above $m_\pi^2$ have been measured, to
serve as an \textsl{ersatz} photon rate measurement
\cite{EXPdilepton}.

The dilepton production rate is indeed related to the real photon
production rate, but they are not quite the same, and we need some
theory input to understand their relation.  At lowest order in
the electromagnetic coupling $\alpha$ and in equilibrium,%
\footnote{%
    In general for a nonequilibrium system, one finds the total (not
    per-unit-volume) rate by replacing $\Pi^<(K)$ with its
    space-integrated version,
    $W^<(K) \equiv \int d^4 X d^4 Y e^{iK\cdot (Y-X)}
    \langle J^\mu(Y) J_\mu(X) \rangle$.  The relationship between
    photon and dilepton production involves whether $k^0=|\k|$
    or $k^0>|\k|$ in the same way.  We will not consider the
    nonequilibrium case because we have so few tools for the calculation
    of fully nonequilibrium correlation functions.
}
both the rate per unit 4-volume to produce a photon and to produce a
dilepton are determined by the current-current correlation
function
\begin{equation}
\label{defPi}
\Pi^<(K) \equiv \int d^4X e^{-i K\cdot X}
    \left\langle J^\mu(0)J_\mu(X)\right\rangle,
\end{equation}
with $J^\mu=eQ\bar\psi\gamma^\mu\psi$ the electromagnetic current operator and with
$\langle .. \rangle$ an average over the quark-gluon plasma ensemble.
In terms of $\Pi^<$, the photon production rate per unit phase space is
\begin{equation}
\label{defratephoton}
\frac{dN_{\gamma}}{d^4 X d^3 \k} \equiv \frac{d\Gamma_{\gamma}}{d^3 \k}
= \left. \frac{1}{(2\pi)^3 2 |\k|}
\Pi^<(K) \right|_{k^0 = |\k|} \,,
\end{equation}
where the photon 4-momentum $K$ is taken on-shell as indicated.  The
dilepton rate exists for any timelike positive-energy $K^\mu$ such as
that $-K^2>4m_l^2$ (with $m_l$ the lepton mass), and away from the
threshold it is given by (see for instance
Ref.~\cite{McLerran:1984ay})
\begin{equation}
\label{defratedilepton}
\frac{d\Gamma_{l\bar l}}{d^4K}=-\frac{2\alpha}{3(2\pi)^4K^2}	
\Pi^<(K) \, \Theta( (k^0)^2 - \k^2) \,.
\end{equation}
The difference, besides the factor of $\alpha/3\pi K^2$, is that the
current-current correlator $\Pi^<(K)$ is evaluated at a timelike
(massive) value for the dilepton rate, and at a null (massless) value
for physical photons.

\begin{figure}[th]
	\begin{center}
	\includegraphics[width=0.7\textwidth]{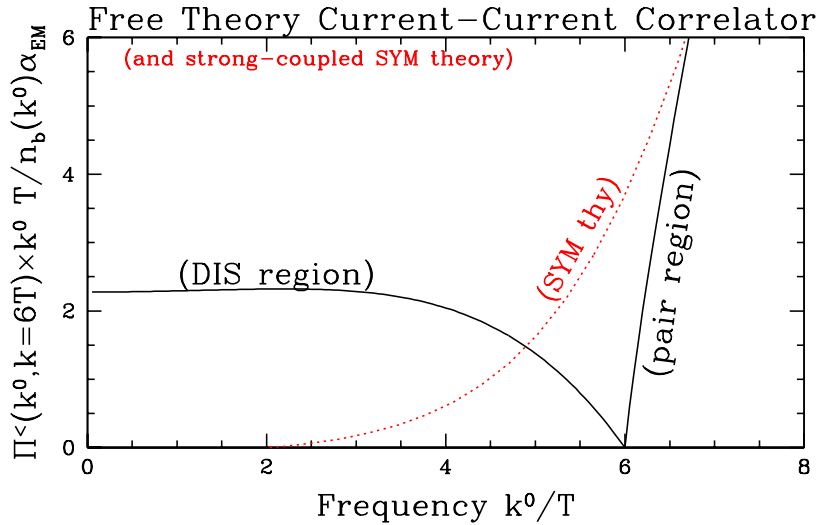}
	\end{center}
\caption{\label{fig_bare}
  The current-current correlator $\Pi^<(K)$ for the case $k=6T$, as a
  function of $k^0$ (normalized by $\alpha$ and its dominant $k^0$
  dependence).  The black curve is for free ($g=0$) QCD, illustrating
  the cusp at $k^0=k$.  The red dotted curve shows the behavior in $N{=}4$ SYM
  theory at infinite coupling, the most strongly coupled QCD-like theory
  known, which shows no cusp-like behavior.}
\end{figure}

So how much does the shift from null to timelike 4-momentum change
things?  \Fig{fig_bare} shows $\Pi^<(K)$ for a fixed
$|\k|=6T$ and a range of $k^0$ values.  The solid (black) curve is a
free-theory level calculation in QCD; the dotted (red) curve is the
value in strongly interacting ${\cal N}{=}4$ SYM theory, from
\cite{CKMSY} (normalized to have the same large-$k^0$ behavior).  In
free (zero-coupling) QCD there is a cusp at the real-photon point.  At
strong coupling, $\Pi^<$ is smooth.  If physical QCD behaves more like
the red curve, then the photon rate and the small-mass-squared dilepton
rate are almost interchangeable.  If it behaves more like the black
curve, then the dilepton rate will show a sharp dependence on the
invariant mass of the dilepton, and photon production will be
suppressed relative to expectations based on moderate invariant-mass
dileptons (if those expectations are based on
\Eq{defratephoton} and \Eq{defratedilepton} and the assumption of smooth
behavior in $\Pi^<$).

The goal of this paper is to provide the most complete perturbative
calculation of $\Pi^<$ for $K^\mu$ close to lightlike which is currently
possible.  Previously, Ref.~\cite{AGMZ} have shown how to compute the
dilepton rate for $K^2$ parametrically in the range $K^2 \sim g^2 T^2$
at leading order in the coupling.  We improve this determination to the
next order in the strong coupling $g$.  We also extend the result to
larger virtuality, $K^2 \sim g T^2$, and discuss the matching onto the
recently completed next-to-leading order calculation at large invariant
mass squared \cite{Laine:2013vma,Ghisoiu:2014mha}.  Our main motivation is to improve
\Fig{fig_bare}, showing how the finite-coupling, perturbative rate
behaves near the real-photon point $K^2=0$.  

Besides the phenomenological justification we have presented, there is
an additional theoretical reason to be interested in doing this.  It is
possible to determine the Euclidean-time-domain behavior of $\Pi$
nonperturbatively on the lattice \cite{latt_people,Ansatzmethod}.
\textsl{At least in principle,} this can be analytically continued to
determine the real-frequency behavior which is physically interesting,
for instance, by applying an Ansatz \cite{Ansatzmethod} or using the
Maximum Entropy Method \cite{MEM}.
Unfortunately, in practice this method is very bad at reconstructing
frequency-domain functions which possess sharp features, such as
that displayed by the black curve in
\Fig{fig_bare}.  This is particularly so if the feature is not
expected and is not built into the model function (priors) used in the
reconstruction.  Therefore, determining whether we expect such a
feature would be very useful in characterizing and improving the
reliability of lattice reconstructions for the photon and dilepton
rates.

The outline of the remainder of the paper is as follows.
We begin in Section~\ref{sec:review} by reviewing the leading-order
calculation of $\Pi^<(K)$ for real photons ($K^2=0$).  We then show the
two extensions in the existing literature:  the extension to small
virtuality, and the extension to next-to-leading order (NLO).
Then, Section~\ref{sec:NLOdilepton} shows how to apply both extensions
at once, giving the NLO dilepton rate at small virtuality.  Next,
Section \ref{sec:results}, presents the results in two ways.  We show
the results in the strict small-$g$ limit.  And we make phenomenological
plots at finite values of $g$.  To do so, we discuss how our calculation
connects onto the recent calculation of Laine and Ghisoiu
\cite{Laine:2013vma,Ghisoiu:2014mha}, which is valid to $\OO(g^2)$
corrections for $-K^2 \sim T^2$.  We end with a discussion, two
technical appendices, and a third appendix collecting tabulated
numerical results.

Very briefly, the two most important take-home results of our work are,
that the dilepton rate at very small invariant mass is increased by
up to $\sim 30\%$ relative to the leading order one, and by a smaller amount
at larger invariant mass.  And we find that high-energy but low-mass
dilepton pairs, such as those recently studied by the PHENIX
collaboration \cite{EXPdilepton}, are probing essentially the same
electromagnetic response as real photons are; so their use as
\textsl{ersatz} photon measurements is valid.

\section{Review of Previous Results}
\label{sec:review}

The real photon rate was first calculated at leading order by Arnold,
Moore, and Yaffe (AMY) in Ref.~\cite{Arnold:2001ba,Arnold:2001ms}.
Aurenche \textsl{et al} showed how to extend this treatment to virtual
photons, that is, dileptons, with small invariant mass, again at leading
order in the coupling \cite{AGMZ}.  Recently, Ghiglieri
\textsl{et al} showed how to extend the AMY treatment of real photons to
the next-to-leading order (NLO) \cite{Ghiglieri:2013gia}.  In this
section we will review the calculations in each of these papers.  The
NLO dilepton calculation will then consist of merging the innovations
in Refs.~\cite{AGMZ,Ghiglieri:2013gia}.

\subsection{Notation}
\label{notation}

Here and throughout the paper capital letters stand for four-vectors, lowercase 
italic letters for the modulus of the spatial three-vectors and the metric signature 
is $({-}{+}{+}{+})$, so that $P^2=p^2-p^2_0$. $K=(k^0,\bk)=(k^0,0,0,k)$ is the  momentum 
of the photon, which we choose to be oriented along the $z$ axis. 
We will work perturbatively in the strong coupling $g$, meaning that we
treat the scale $gT$ (the soft scale) as parametrically smaller than the
scale $T$ (the hard scale).

Throughout the paper we will often use light-cone coordinates, which we
define as $p^{-}\equiv p^0- p^z$ and $p^+ \equiv \frac{p^0 + p^z}{2}$.
This normalization is nonstandard, but we find it convenient because
$dp^0 dp^z = dp^+ dp^-$, and because we will frequently encounter cases
in which $p^-=0$, in which case $p^z = p^0 = p^+$ with our conventions.
The transverse coordinates are written as $\bpp$, with modulus $\pp$.
With this choice, we will treat the plus component of the photon momentum, $k^+$,
to be of the order of the temperature or larger, $k^+\simg T$, while the other component
is assumed to be of order $g^2T$, $k^-\sim g^2T$, so that $-K^2=2k^+k^-\sim g^2T^2$.

\subsection{Leading order Photons}
\label{sec:LOphoton}

\begin{figure}[ht]
	\begin{center}
	\includegraphics[width=9cm]{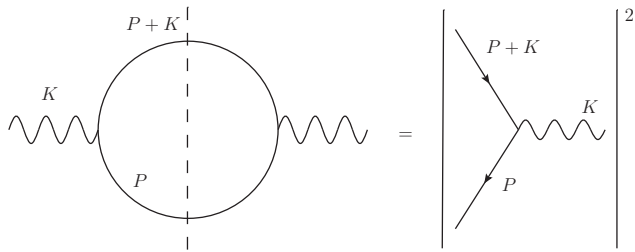}
	\end{center}
	\caption{The Born diagram on the left. Its cut is associated
          with the square of the thermal Drell-Yan process on the
          right.  This process is kinematically forbidden for a real
          photon and on-shell quarks.}
	\label{fig_born}
\end{figure}

\begin{figure}[ht]
	\begin{center}
	\includegraphics[width=12.85cm]{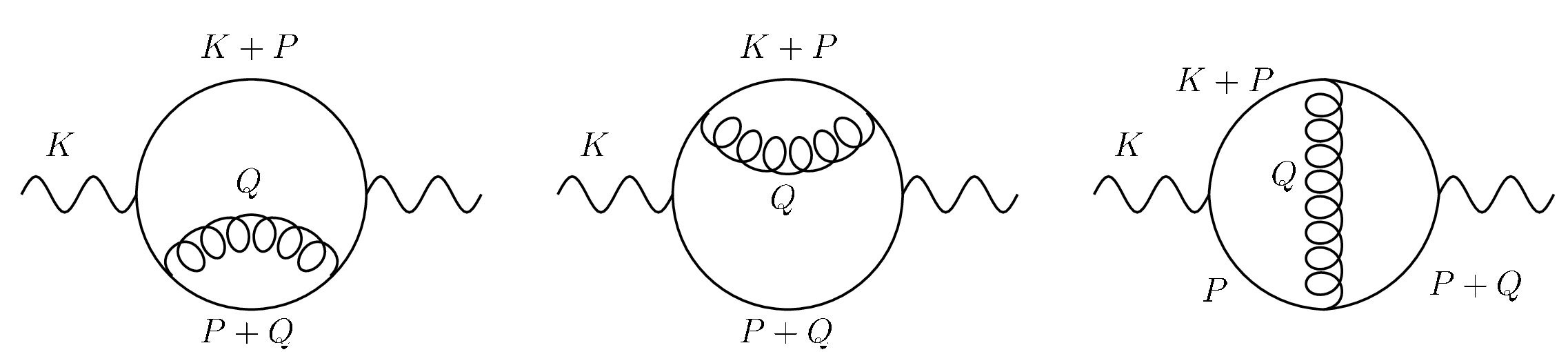}
	\end{center}
	\caption{Two-loop diagrams. Curly lines are gluons. 
	The momentum assignments shown here will be used throughout the paper.}
	\label{fig_diagrams}
\end{figure}

In a plasma without net baryon number%
\footnote{For the case with a net baryon number there are additional
  complications, see \cite{GaleMajumder,Gervais:2012wd}.  We will not consider this
  case in the present paper.},
the photon self-energy $\Pi^<(K)$ is determined by diagrams with two
current insertions (which we represent as external photon lines)
attached to a quark loop, with an arbitrary number of gluonic
corrections.  AMY have shown \cite{Arnold:2001ba} that only some of
these diagrams contribute, and only in some kinematical regions.  In
particular, the lowest-order diagram, shown in \Fig{fig_born},
corresponds to a kinematically disallowed process.  Instead the lowest
order diagrams involve at least one gluon line, as depicted in
\Fig{fig_diagrams}.  The figures also establish the momentum
notation we will use throughout:  the quark lines have momentum $P$
and $P\pm K$, while gluon lines have momentum $Q$.  They also show that
the rate will depend parametrically on couplings as $e^2 g^2$.

\begin{figure}[ht]
\begin{center}
\includegraphics[width=0.8\textwidth]{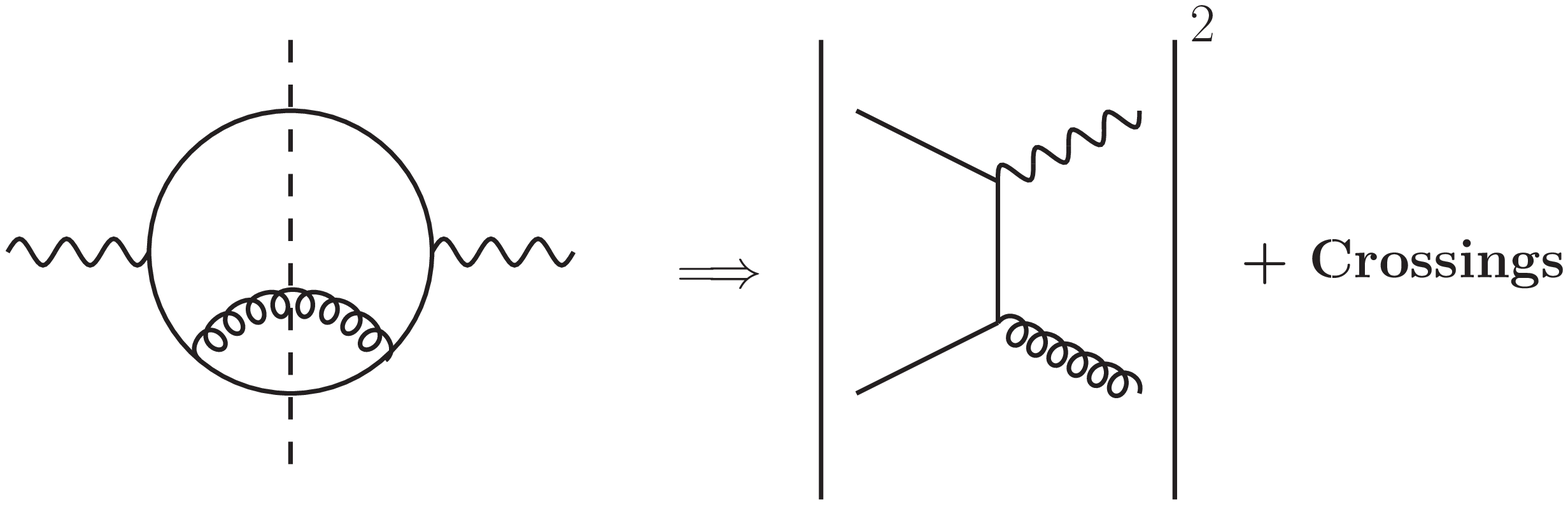}
\end{center}
\caption{\label{fig_hard_cuts}
 Cut of a two loop diagram (left) corresponds to a $\twotwo$ scattering
 process (right).}
\end{figure}

\begin{figure}
\begin{center}
\includegraphics[width=0.8\textwidth]{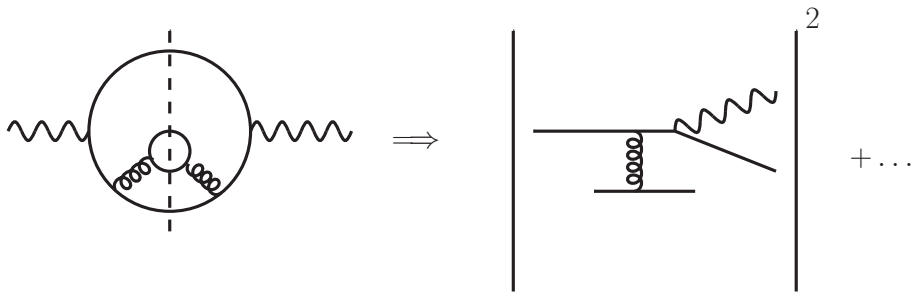}
\end{center}
\caption{\label{fig_scatt}
  Two-loop diagram cut through a self-energy correction on the gluon,
  which corresponds to scattering-induced photon radiation (crossings
  not shown)}
\end{figure}

\begin{figure}
\begin{center}
\includegraphics[width=0.75\textwidth]{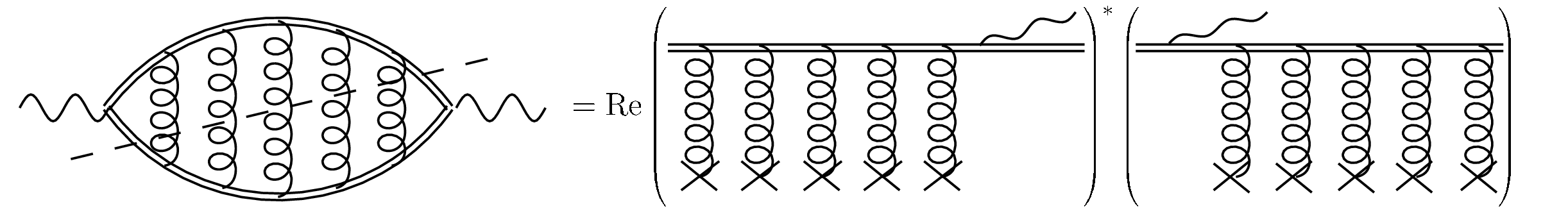}
\end{center}
\caption{\label{fig_lpm}
  Example of how a cut ``multi-rung'' diagram corresponds to
  interference between processes where a photon is emitted before or
  after a series of scatterings (gluons connecting to $X$'s).  Such
  diagrams must be resummed to determine the leading-order photon
  production rate in the collinear kinematic range.}
\end{figure}

In evaluating the cut diagrams in \Fig{fig_diagrams}, there are
contributions where the gluon line is an on-shell external gluon
state.  The resulting processes, shown in \Fig{fig_hard_cuts},
are leading-order for quark momentum in the
$p \in [gT,T]$ momentum range, and were traditionally assumed to give
the leading-order photon production rate
\cite{Kapusta:1991qp,Baier:1991em}.  We will call these $\twotwo$
processes or elastic processes.  But another possibility is that the
gluon line can be soft, $q \sim gT$, and spacelike off-shell.  In this
case it is essential to include the Hard Thermal Loop (HTL)
resummations to the gluon propagator \cite{Braaten,Frenkel}, under
which the gluon has significant spectral weight in this region.  This
leads to a distinct contributing kinematical region which
corresponds to scattering-induced emission, as shown in
\Fig{fig_scatt}.  We will call these collinear processes or collinear
splitting processes.  Aurenche \textsl{et al} \cite{Aurenche:1998nw}
first showed that these processes are also leading order and can even
be numerically dominant.  The reason is that the process includes a
kinematical region in which the intermediate quark line in
\Fig{fig_scatt} is nearly on the mass shell.  But this
near-singularity requires the inclusion of self-energy corrections,
which bring in additional diagrams by gauge invariance and the
necessity to correctly represent charge conservation.  Therefore, in
the kinematic region where gluons are soft and spacelike (representing
scattering processes), one must sum over multiple gluon exchanges,
such as the diagram of \Fig{fig_lpm}. The interference effect this
generates and the associated suppression are called the Landau-Pomeranchuk-Migdal
(LPM) effect.

\begin{figure}
\begin{center}
\includegraphics[width=0.5\textwidth]{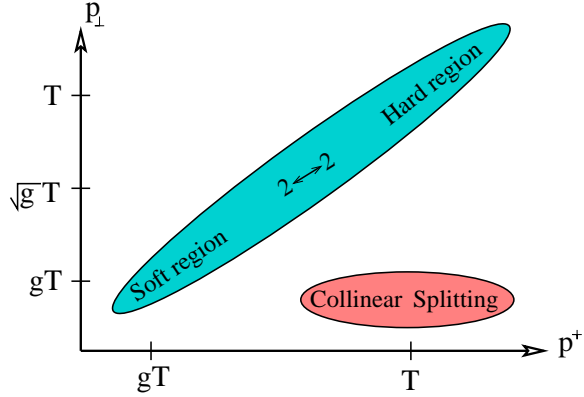}
\end{center}
\caption{\label{lomap}
  Distinct momentum regions giving rise to leading-order photon
  production, in terms of the off-shell fermionic momentum components
  $\bpp$ and $p^+$.  The $\twotwo$ processes occur when these are
  comparable, the collinear processes occur when $p^+ \gg \pp$.
}
\end{figure}

In \cite{Arnold:2001ba}, AMY showed that these two kinds of processes
(elastic scattering when one gluon is on-shell, scattering induced
emission with any number of soft spacelike gluons) are both needed in
the calculation, but arise from kinematically distinct momentum
regions.  Therefore the computation can be separated into a
contribution from each process.  The easiest way to see that this is
true is to consider the components of the off-shell fermion's momentum
$P$, particularly the transverse component $\bpp$ and the longitudinal
component $p^+$.  As illustrated in \Fig{lomap}, the relevant momentum
regions are quite distinct when viewed with this variable.

The required diagrams in
the collinear splitting region can be resummed into an integral equation for the
production rate.  The analysis is quite detailed, and rather than
reproduce it here, we will just quote the key results.

The contribution to $\Pi^<(K)$, for $K^2=0$, from elastic scattering and
at lowest order in the strong coupling, arises from the scattering
processes
$q_{k'} g_{l'} \to \gamma_k q_l$
 and
$q_{k'} \bar{q}_{l'} \to \gamma_k g_l$
(and processes with
$q \leftrightarrow \bar{q}$), which contribute
(writing $k^0=|\k|=k$)
\begin{eqnarray}
\label{Pi22LO}
\Pi_{\twotwo,\sss{LO}}^<(K) & = & \frac{48\pi \Bb(k)}{T^2}
\int \frac{d^3 \k' d^3 \ell' d^3 \ell}{(2\pi)^9 8l l' k'} 
 (2\pi)^4 \delta^4(K'_\mu{+}L'_{\mu}-K_{\mu}-L_{\mu}) \times
 \\ && \hspace{2em}
\times \left( \frac{\nfd(k') \nbe(l')(1{-}\nfd(l))}{\nfd(k)}
\left[ \frac{-s}{t} - \frac{t}{s} \right]
 + \frac{\nfd(k') \nfd(l') (1{+}\nbe(l))}{\nfd(k)}
\left[ \frac{u}{t} \right] \right),   \nn
\end{eqnarray}
where $s,t,u$ are the usual Mandelstam variables and
\begin{equation}
	\label{B_of_k}
	\Bb(k) =  \alpha_\mathrm{EM} \crr g^2 T^2 \nfd(k)
	  \sum_s \dr q_s^2
\quad \left( = \frac{8 \alpha_\mathrm{EM} \nfd(k) g^2 T^2}{3 }  \, 
\mbox{for QCD with $uds$ quarks} \right)
\end{equation}
is the leading-log coefficient, which we will use to normalize all
contributions to $\Pi^<(K)$ in what follows.
This expression is based on massless dispersion for the external states
and vacuum matrix elements.  

\hspace{-0.4em}
To evaluate this expression it is convenient to work in terms of the
transfer 4-momentum $P_\mu \equiv K'_\mu - K_\mu$, which is the momentum
carried by the intermediate off-shell quark line in diagrams giving rise
to the $1/t$ matrix element.  For small $p\sim gT$ the on-shell condition
$(K')^2=0$ enforces $p^-=\OO(g^2 T)$, and we may simplify $-s/t$, $u/t$,
and the statistical functions enough to perform all integrations but
the $p^+,\pp$ integrals:
\begin{equation}
\label{softedge}
\Pi^<_{\twotwo,\sss{LO}}(K)
 \; \stackrel{\mathrm{small}\:p}{\longrightarrow}\;
\Bb \int_0 \pp d\pp \int_{-\infty}^{\infty} dp^+ 
  \frac{1}{2(\pp^2+p^+{}^2)^{3/2}} = \Bb \int_0 \frac{d\pp}{\pp},
\end{equation}
which is log divergent.  The divergence is removed because for small
$p\sim gT$ there are HTL \cite{Braaten,Frenkel} medium corrections to
the fermionic self-energy which change the matrix elements.
We showed in \cite{Ghiglieri:2013gia}
that, by working in the $p_\perp,p^+$ basis, these modifications can be
handled analytically and the $p^+$ integral can still be performed; the
infrared limit of the $\pp$ integral is modified to
\begin{equation}
\label{sumrule}
\Pi^<_{\twotwo,\sss{LO}}(K)
 \; \stackrel{\mathrm{small}\:p}{\longrightarrow}\;
\Bb \int_0 \frac{\pp d\pp}{\pp^2 + \mm} \,,
\end{equation}
where $\mm$ is the thermal asymptotic%
\footnote{
    $\mm = 2 m_q^2$ the ``mass'' of a quark at rest, which is used in
  much of the HTL literature \cite{Braaten}.    
    }
quark mass $\mm = g^2 T^2 \crr/4$.
Matching this behavior with the large-$p$ behavior which is found
numerically, one finds
\begin{eqnarray}
\Pi_{\twotwo,\sss{LO}}^<(K) & = & \Bb \left[ \ln\left( \frac{T}{\justm}\right)
  + C_\mathrm{hard}\left( \frac{k}{T} \right) \right] \,,
\nn \\
C_\mathrm{\hard}\left(\frac{k}{T}\right) & \approx & 
  \frac12 \ln\left(\frac{2k}{T}\right)
 + 0.041\frac{T}{k} -0.3615 + 1.01 e^{-1.35 k/T},
 \quad 0.2<\frac{k}{T} .
\label{Chard}
\end{eqnarray}
The result for $C_\mathrm{hard}$ is a numerical fit.

\begin{figure}[ht]
\begin{center}
\includegraphics[width=7cm]{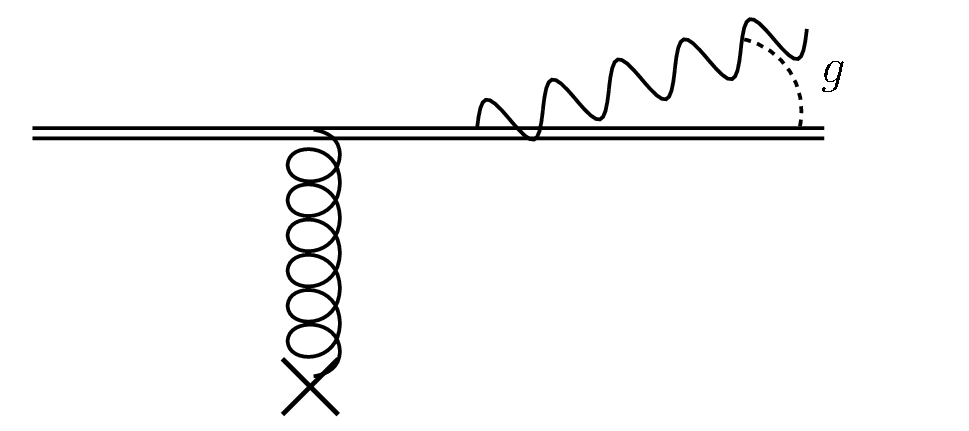}
\includegraphics[width=6.7cm]{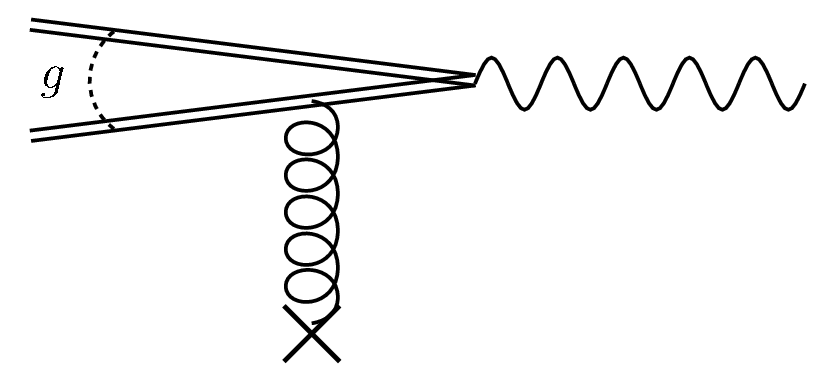}
\end{center}
\caption{Collinear diagrams.
  The diagrams where the  gluon is attached to the other fermionic line
  are not shown.  In both graphs the gluon is soft and is scattering off
  the hard quarks and gluons  of the plasma as indicated by the crosses,
  \emph{i.e.}, it is an HTL gluon in the Landau cut.
}
\label{fig_collinear}
\end{figure}

Next it is necessary to consider interactions with one or more spacelike
off-shell gluons (that is, gluon-exchange scatterings).  An efficient
way to think of these processes is that the scattering provides a little
off-shellness such that the process of \Fig{fig_born} becomes
kinematically allowed.  This is only efficient for a narrow range of
transverse momenta $\pp^2 \sim g^2 T^2$, leading to a $g^2$ phase space
suppression.  The relevant diagrams are shown in
\Fig{fig_collinear}, which emphasizes the small angular spread
between the photon and the emitting particle.  It is because this angle is
narrow that we refer to these as collinear processes.

The deviation from the mass shell is $\delta p^0 \sim g^2 T$, which
allows for large off-shell propagation distances $d\sim 1/g^2 T$, which
is the same order as the mean inter-scattering distance.  Therefore one
must resum multiple scattering, as illustrated in \Fig{fig_lpm}.
The conceptual issues involved in such a resummation are treated in
\cite{Zakharov} and the technical issues are in \cite{Arnold:2001ba}.
Skipping the details, the current correlator from these processes, for
on-shell 4-momentum $K$ and at leading order, is given by
\begin{eqnarray}
\Pi_{\coll}^<(K) & = & \Bb
	\int_{-\infty}^\infty dp^+ 
        \frac{\nfd(k{+}p^+) [1-\nfd(p^+)]}{\nfd(k)}
	\nn \\
	\label{fpprob1}
	 & & \times \frac{1}{\mm} \int \frac{d^2\pp}{(2\pi)^2}
	  \:{\mathrm Re}\left[ \frac{(p^+)^2 + (p^+{+}k)^2}{2(p^+)^2 (p^+{+}k)^2}
	 \bpp \cdot \f(\bpp) \right] ,
\\
\label{qspace1}
 2\bpp & = & i\delta E \; \f(\bpp) +
 \int \frac{d^2\qp}{(2\pi)^2} \: \cc(\qp)
  \Big[ \f(\bpp) - \f(\bpp {+} \q_\perp) \Big] \,,
\\
\label{deltaE1}
\delta E & = & \frac{k(\pp^2 + \mm)}{2p^+(k{+}p^+)} \,.
\end{eqnarray}
Here $\delta E=E_{\bp+\bk}-E_\bp-k^0$ is the eikonalized energy
difference between having a quark of energy corresponding to a momentum
of $k+p$ and having a quark of energy corresponding to momentum $p$ with
a photon of energy and momentum $k$. $\cc(\qp)$ is the differential soft
scattering rate, which at leading order reads
\cite{Arnold:2001ba,Aurenche:2002pd}
\begin{equation}
\label{C_LO}
 \cc(\qp) =g^2\crr\int\frac{dq^+dq^-}{(2\pi)^2}2\pi\delta(q^-)
 G^{rr}_{\mu\nu}(Q)v_k^\mu v_k^\nu = 
     g^2\crr T \frac{\md^2}{\qp^2(\qp^2{+}\md^2)} \,,
\end{equation}
where $v_k^\mu\equiv(1,0,0,1)$ and $G_{\mu \nu}^{rr}$ is the cut
(symmetric) HTL gluon propagator.  Here $\md$ is the Debye mass,
$\md^2=g^2T^2(\ca + \nf T_R)/3 = g^2 T^2(1+\nf/6)$ at leading order.

\Eq{fpprob1} can be understood in the following way. Real
photon production involves a current operator insertion in the amplitude,
followed by time evolution, and then a current insertion in the
conjugate amplitude.  At times between the current insertions, the
density matrix contains an off-diagonal term with a quark with momentum
$p^+,\bpp$ and a photon of momentum $K$ in the amplitude, but a
quark of momentum $(p^+{+}k),\bpp$ in the conjugate amplitude.
The size for this entry in the density matrix is $\f(\bpp)$ and \Eq{qspace1}
is the time-integrated evolution equation for this amplitude;
$\delta E$ represents the phase accumulation because the state in the
amplitude and conjugate amplitude do not have the same energy, while the
$\cc(\qp)$ term describes the effect of scattering processes on the
evolution of the density matrix.  The second line in \Eq{fpprob1}
describes the overlap of the current operator on this density matrix
element, times the DGLAP splitting kernel.

\Eq{qspace1} and \Eq{fpprob1} were solved explicitly in
Ref.~\cite{Arnold:2001ms}.  We will not present the details here, since
we will have to solve a modified version of this relation.

\subsection{Leading order Dileptons}
\label{sec:LOdilepton}

Now let us look at the modifications necessary to consider slightly
timelike $K^\mu$, that is, $k^+ = \frac{k^0+k}{2} \sim T$ as before, but 
$k^- = (k^0 - k) \sim g^2 T$ rather than strictly $k^-=0$.  This
frequency difference is of order the characteristic soft plasma scattering
rate, meaning that no particle in the plasma is on-shell to better
precision than $\OO(g^2 T)$.  Therefore such a small virtuality does not
change any of the parametric analyses, and the set of diagrams is
unchanged.  But we must re-consider their evaluations.

First consider the $\twotwo$ processes.  The processes shown in
\Fig{fig_hard_cuts} always involve an off-shell intermediate line.
Because of plasma screening, the leading-order contribution only arises
when this line's momentum is $\gsim gT$.  Therefore the leading-order
$\twotwo$ results are not changed by $k^- \sim \OO(g^2 T)$.

The same is not true of collinear processes, because the virtuality
involved is $\delta E \sim g^2 T$.  Therefore the introduction of
$k^- \sim g^2 T$ makes an important difference, even making the
tree-level (Born) process kinematically allowed if $k^-$ is large
enough.  Aurenche \textsl{et al} present a complete leading-order
analysis \cite{AGMZ} (see also \cite{Carrington:2007gt}), showing that \Eq{fpprob1} through \Eq{deltaE1} are
changed to
\begin{eqnarray}
\Pi_{\coll}^<(K) & = & \Bb
	\int_{-\infty}^\infty dp^+ 
        \frac{\nfd(k^+{+}p^+) [1-\nfd(p^+)]}{\nfd(k^+)}
	\nn \\
	\label{fpprob2}
	 & & \times \frac{1}{\mm} \int \frac{d^2\pp}{(2\pi)^2}
	  \:{\mathrm Re}\left[ \frac{(p^+)^2 + (p^+{+}k^+)^2}
            {2(p^+)^2 (p^+{+}k^+)^2}
	 \bpp \cdot \f(\bpp) +\frac{2k^-}{k^+}g(\bpp) \right] ,
\\
\label{qspace2}
 2\bpp & = & i\delta E \; \f(\bpp) +
 \int \frac{d^2\qp}{(2\pi)^2} \: \cc(\qp)
  \Big[ \f(\bpp) - \f(\bpp {+} \q_\perp) \Big] \,,
\\
\label{migdallong}
2	&=&i\delta E\; g(\bpp)+
 \int \frac{d^2\qp}{(2\pi)^2} \: \cc(\qp)
  \Big[ g(\bpp) - g(\bpp {+} \q_\perp) \Big]\,,
\\
\label{deltaE2}
\delta E & = & \frac{k^+(\pp^2 + \mm)}{2p^+(k^+{+}p^+)} + k^- \,.
\end{eqnarray}
The changes are that, first, $\delta E$ receives an extra contribution
from $k^-$, and, second, there is a new contribution $g(\bpp)$
accounting for the contribution of longitudinal photons, which are
absent for $k^-=0$.

We should remark that these expressions are also valid for $k^- < 0$,
that is, spacelike $K^2$, which corresponds physically to probing
the plasma through deep inelastic scattering.  In this regime the
longitudinal term $g(\bpp)$ corresponds to a timelike photon exchange
and gives a negative contribution.  While the DIS regime is not
experimentally accessible, we still study it, because the continuity
or smoothness of $\Pi^<$ is relevant to the lattice reconstruction of
the spectral function.

\subsection{Real photons at NLO}
\label{sec:NLOphoton}

Now we return to real photons, $k^- = 0$, but we consider corrections at
the next order in the strong coupling $g$.  This case has been
considered in detail in Ref.~\cite{Ghiglieri:2013gia}, so again we will
summarize the results without presenting a detailed derivation.

In vacuum field theory, NLO corrections generally arise from a loop
or an extra external leg and
are suppressed by $\OO(g^2)$.  In thermal field theory,
NLO corrections can be as large as $\OO(g)$ whenever soft $p\sim gT$
momenta are involved, since bosonic lines with $\OO(gT)$ momenta can
introduce large Bose factors $\nbe(p^0) \sim T/p^0 \sim 1/g$ when
$p^0\sim gT$.  Therefore we must examine our calculation for such soft
sensitivity.

\begin{figure}
\begin{center}
\includegraphics[width=0.6\textwidth]{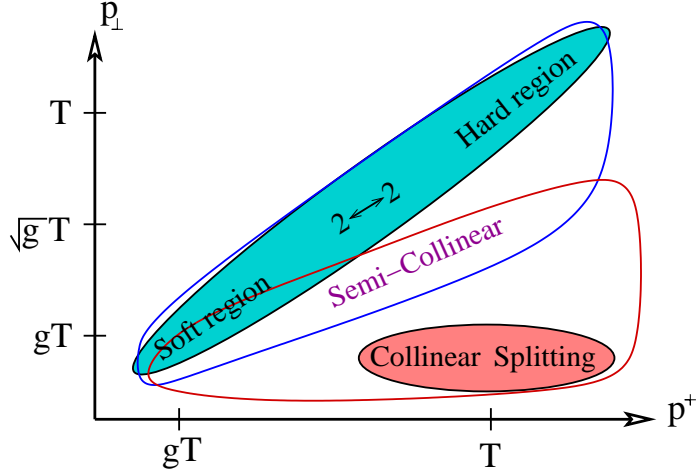}
\end{center}
\caption{\label{nlomap}
  Phase space regions in the $\pp$, $p^+$ plane which are relevant at
  NLO.  The collinear region expands towards small $p^+$ and larger
  $\pp$, while the $\twotwo$ region expands to smaller $\pp$.  They
  overlap in the semi-collinear region.}
\end{figure}

The $\twotwo$ calculation receives $\OO(g)$ corrections because an
$\OO(g)$ fraction of the contribution in \Eq{Pi22LO} arises when the
argument of $\nbe(l)$ or $\nbe(l')$ is $\OO(gT)$.  In this region our
treatment of the external state dispersion and spectral weight break
down, so we must reconsider this $\OO(g)$ contribution.  In this region
$s,t\sim gT^2$, so the other momentum combinations are at a relatively
small opening angle $\theta \sim \sqrt{g}$.  Therefore we call this a
semi-collinear contribution.  There can also be $\OO(g)$ corrections
more generally when the exchanged fermion becomes soft, $t\sim g^2
T^2$. We illustrate this expansion of the relevant $P$-momentum region
in \Fig{nlomap}:  while at leading order we need the blue shaded region,
at NLO we need the region enclosed in the larger blue contour.

The collinear calculation has $\OO(g)$ corrections from several places.
First, the ingredients in \Eq{qspace1} receive $\OO(g)$ corrections,
\begin{equation}
\cc(\qp) \to \cc(\qp) + \delta \cc(\qp) \,,
\qquad
\mm \to \mm + \delta \mm \,,
\label{dC,dm}
\end{equation}
where $\delta \mm = -2\md \mm / (\pi T)$ \cite{CaronHuot:2008uw} and
the explicit expression for $\delta \cc$ can be found in
Ref.~\cite{CaronHuot:2008ni}; both effects are summarized in
\cite{Ghiglieri:2013gia} Eq.~(B.27) and Eq.~(C.1).  In addition, an
$\OO(g)$ portion of the contribution arises from $p^+\sim gT$ and an
$\OO(g)$ portion arises from $\pp^2 \sim gT^2$.  In each region at least
one collinear approximation used to derive \Eq{fpprob1}, \Eq{qspace1}, or
\Eq{deltaE1} breaks down -- though in the $\pp^2 \sim g^2 T^2$ region
the collinear approximations only suffer $\OO(g^2)$ correction.  The
expanded phase space region relevant at
NLO is again illustrated in \Fig{nlomap} as the larger red contour.
Note that this contour overlaps with the blue ($\twotwo$) contour in the
soft region and in the semi-collinear region with $p^+\sim T$ and
$\pp\sim \sqrt{g}\,T$.  These regions will need special attention.

We can handle the corrections $\delta \mm$ and $\delta \cc$ as follows.
We will replace $\f(\bpp) \to \f(\bpp) + \delta\f(\bpp)$, with
$\delta \f$ the $\OO(g)$ correction arising at linear order from
$\delta \mm$ and $\delta \cc$.  Specifically, \Eq{qspace1} becomes
\begin{equation}
\label{qspaceNLO}
2\bpp = i (\delta E + \delta E')(\f+\delta \f)
+ \int \frac{d^2 \qp}{(2\pi)^2} ( \cc + \delta \cc )
\Big[ \f(\bpp) + \delta f(\bpp) - \f(\bpp+\bqp)
 - \delta f(\bpp+\bqp) \Big],
\end{equation}
with $\delta E' = \frac{k \delta \mm}{2p^+(k{+}p^+)}$.  Expanding at
zero order in $\delta \mm$ and $\delta \cc$ we recover \Eq{qspace1},
while at first order we find
\begin{eqnarray}
\label{df_NLO}
&& i \delta E \delta \f(\bpp)
+ \int \frac{d^2 \qp}{(2\pi)^2} \cc(\qp)
\Big[ \delta \f(\bpp)  - \delta \f(\bpp+\bqp) \Big]
\nn \\ & = &
-i \delta E' \f(\bpp)
- \int \frac{d^2 \qp}{(2\pi)^2} \delta \cc(\qp)
\Big[ \f(\bpp)  - \f(\bpp+\bqp) \Big] \,.
\end{eqnarray}
Once $\f(\bpp)$ has been determined, it acts as a source in an
inhomogeneous equation for $\delta f(\bpp)$.  We present a particularly
clean methodology to treat these integral equations in Appendix
\ref{app_hankel}, but we do not present the
results here because they will be superseded in the next section.  The
corrections remain subdominant in the large-$\pp$ and small-$p^+$
regions, so there is no overlap between these corrections and those
from the kinematic edges of the collinear region.

Treating the kinematic edges, that is, the soft and semi-collinear
regions, turns out to be surprisingly easy.  The solution is to perform
the leading-order calculations as if nothing has changed, to evaluate
the NLO collinear corrections from $\delta\mm$ and $\delta \cc$,
and to add the following next-to-leading correction from the overlap
regions:
\begin{equation}
\label{Csoft+sc}
\delta\Pi^<_\mathrm{NLO,sc}(k) = \Bb \frac{-\md}{\pi T}
\left( \ln \frac{2\md T}{\mm} + 2 C_\mathrm{soft+sc}\left(\frac{k}{T}\right) \right),
\end{equation}
where $C_\mathrm{soft+sc}(k)$ is given by Eq.(6.3) of
Ref.~\cite{Ghiglieri:2013gia}.

\section{Dileptons at NLO}
\label{sec:NLOdilepton}

In the previous section we saw how to extend the leading-order photon
production calculation of Arnold, Moore, and Yaffe in two ways.  We
extended it to nonzero but small photon virtuality; and we extended it
to next-to-leading order.  Now we want to manage both extensions
simultaneously.  To do this, we need to figure out how the NLO
corrections detailed in Subsection \ref{sec:NLOphoton} get modified by
the virtuality discussed in Subsection \ref{sec:LOdilepton}.

\subsection{NLO effect of virtuality}
\label{sec:NLOvirt}

A key step in the leading-order dilepton treatment was showing that the
$\twotwo$ processes are not disturbed at leading order by the
introduction of a small photon virtuality $k^- \sim g^2 T$.  We now show that this
persists to NLO (defined as $\OO(g)$).  To do so, we will have to look
momentum region by momentum region.  The hard scattering region
(\Eq{Pi22LO} with $l,l',k'\sim T$) is the simplest to analyze.  An
$\OO(g^2 T)$ shift to $k^0$ changes the statistical functions, the
matrix element, and the phase space only by $\OO(g^2)$ amounts, and can
therefore be neglected.  The semi-collinear region is also
straightforward.  If we consider the $u/t$ term and $l\sim gT$, we find
that the shift to $k^0$ modifies the value of $l$ by an $\OO(g^2 T)$
amount.  This will change the phase space, $\nbe(k')$, and $u/t$ by
$\OO(g)$ amounts.  But since this region is already only an $\OO(g)$
correction, the impact of these shifts is $\OO(g^2)$.  We can also see
this by considering the semi-collinear region from the collinear side.
If we consider \Eq{deltaE2} in the region $\pp^2\sim
gT^2$, an $\OO(g^2 T)$ value for $k^-$ is an $\OO(g)$ correction to the
$\pp^2$ term.  And the contribution in \Eq{fpprob2} from $g(\bpp)$
is $\OO(g)$ relative to the contribution from $\f(\bpp)$, due to the
$\bpp$ factors associated with the latter.  Therefore, in this kinematic
region we again find that $k^- \sim g^2 T$ leads to an $\OO(g)$
correction to an already $\OO(g)$ suppressed region, and can be
neglected.

The analysis of the soft region is more subtle.  We must reconsider the
approximations leading to \Eq{softedge}, allowing $k^-\sim g^2 T$.
The shift has only an $\OO(g^2)$ bearing on the statistical functions
and $\ell,\ell'$ integrations; but it will shift the value of $P^\mu$.
Specifically, the on-shell condition $(K')^2=-\mm$ means that
$p^-$ shifts from $\frac{\mm+\pp^2}{2k^+}$ to
$\frac{\mm+\pp^2}{2k^+} - k^-$.  In deriving \Eq{softedge},
\Eq{sumrule}, we assumed that $p^- = 0$; while in fact it is smaller
than $\pp$ by a relative factor of $g$.  But we showed
in Ref.~\cite{Ghiglieri:2013gia} that finite-$p^-$ corrections to
\Eq{sumrule} only arise at even orders in $p^-$.  Therefore a
relative-order $g$ suppressed correction to $p^-$ leads to an $\OO(g^2)$
correction, which can be neglected.
Hence the soft and semi-collinear corrections, summarized in
\Eq{Csoft+sc}, are not modified at $\OO(g)$ order.

So the only place where $k^-\!\sim g^2 T$ can affect $\Pi^<$ at
$\OO(g)$ is in the collinear phase space region.  The collinear
expansion which establishes \Eq{fpprob2}, \Eq{qspace2}, \Eq{migdallong},
and \Eq{deltaE2} is an expansion in $k^- \ll T$, $p_\perp^2 \ll T^2$, so
a small $\OO(g^2 T)$ correction will only modify the \textsl{structure}
of these equations at the $\OO(g^2)$ level.  So the only modifications
we need at the NLO level are the same ones we needed for real photons;
the NLO corrections $\delta \cc$ and $\delta \mm$.

Hence, an NLO calculation of the dilepton rate requires the NLO
treatment of real photons from $\twotwo$ and semi-collinear regions, and
a treatment of \Eq{fpprob2} to \Eq{deltaE2} but including NLO
corrections using \Eq{df_NLO} and the analogous expression for
$\delta g(\bpp)$,
\begin{eqnarray}
\label{dg_NLO}
&& i \delta E \delta g(\bpp)
+ \int \frac{d^2 \qp}{(2\pi)^2} \cc(\qp)
\Big[ \delta g(\bpp)  - \delta g(\bpp+\bqp) \Big]
\nn \\ & = &
-i \delta E' g(\bpp)
- \int \frac{d^2 \qp}{(2\pi)^2} \delta \cc(\qp)
\Big[ g(\bpp)  - g(\bpp+\bqp) \Big] \,.
\end{eqnarray}
We specify our numerical procedure in
Appendix~\ref{app_hankel}.

\subsection{Emergence of the Born term}
\label{sec:Born}

As shown in \cite{AGMZ}, \Eq{fpprob2} includes the Born
term, \textsl{i.e.}, the contribution from the thermal Drell-Yan
process show in \Fig{fig_born}.  Let us explore how this occurs, and use it to study how the
solutions behave as $k^-$ becomes larger, in particular for
$k^+ k^- \gg \mm$.  If the collision operator $\int d^2\qp
\cc(\qp)[\f(\bpp)-\f(\bpp+\bqp)]$ is treated as formally much smaller
than $\delta E$,  Eqs.~\eqref{qspace2} and \eqref{migdallong} can be
solved by substitution order by order in the collision operator.
The zeroth order, $\f^{(0)}(\bpp)$ and $g^{(0)}(\bpp)$, corresponds
to having no exchanged soft gluons.
In more detail one has for the transverse function
\begin{equation}
	\label{zerothtransverse}
	 2\bpp  =  i\delta E \; \f^{(0)}(\bpp)\,, 
\end{equation}
so that
\begin{equation}
	\label{transverseborn}
  \int\frac{d^2\pp}{(2\pi)}\mathrm{Re}\,\bpp\cdot\f^{(0)}(\bpp)
  =\int\frac{d^2\pp}{(2\pi)}\mathrm{Re} \frac{-i2\pp^2}
  {\frac{k^+(\pp^2 + \mm)}{2p^+(k^+{+}p^+)}+k^--i\epsilon}  ,
\end{equation}
where we have included an $i\epsilon$ prescription to account for a very
small contribution from $\cc$.  The real part only arises due to the
$i\epsilon$ prescription, and is a delta function at
the point where the denominator vanishes.  One then obtains
\begin{equation}
 \int\!\!\frac{d^2\pp}{(2\pi)^2}\mathrm{Re}\,\bpp\cdot\f(\bpp)
  = \frac{\left\vert p^+(k^+{+}p^+)\right\vert}{k^+}
 \left(\frac{p^+(k^+{+}p^+)}{(k^+)^2}K^2{-}\mm\right)
\theta\left(\frac{p^+(k^+{+}p^+)}{(k^+)^2}K^2{-}\mm\right).
	\label{finaltransverseborn}
\end{equation}
The first term is the coefficient on $\pp^2$ in the denominator of
\Eq{transverseborn}, the second term is the value of $\pp^2$ in the
numerator, and the Heaviside function makes sure to consider only cases
where the denominator vanishes for positive $\pp^2$.
The longitudinal mode is evaluated in the same fashion, leading to a
complete Born term of
\cite{Aurenche:2002pc,AGMZ}
\begin{equation}
 \Pi_\mathrm{coll,Born}^<(K) = 
%\frac{\Bb}{\nfd(k^+)}
\Bb
	\int_{p^+_\mathrm{min}}^{p^+_\mathrm{max}} dp^+
%        \nfd(k^+{+}p^+) [1 {-} \nfd(p^+)]
        \frac{\nfd(k^+{+}p^+) [1 {-} \nfd(p^+)]}{\nfd(k^+)}
	\: \left[ \frac{k^-}{\mm}
+\frac{(p^+)^2 + (p^+{+}k^+)^2}{2k^+\,p^+\, (p^+{+}k^+)}
	 \right],
\label{fpborntot}
\end{equation}
where $p^+_\mathrm{min}=-\frac{k^+}{2}(1+\sqrt{1+4\mm/K^2}\,)$,
$p^+_\mathrm{max}=-\frac{k^+}{2}(1-\sqrt{1+4\mm/K^2}\,)$.
Because the quarks have (effective) mass $\mm$, the Born term is only
kinematically allowed for $-K^2>4\mm$, as reflected by these
integration limits.

Next consider $k^0 < 0$ (the DIS regime).  \Eq{finaltransverseborn} is
valid for either sign of $k^-$, but the Heaviside function picks out a
different range of $p^+$, namely
$p^+ > p^+_\mathrm{DIS}=k^+/2(\sqrt{1+4\mm/K^2}-1)$
(and $p^+ < -k-p^+_\mathrm{DIS}$, which is symmetrical).  Then
\Eq{fpborntot} is replaced with
\begin{equation}
	\Pi^<_\mathrm{coll,Born,DIS}(K^2) = 
%\frac{\Bb}{\nfd(k^+)} 
\Bb
	\int_{p^+_\mathrm{DIS}}^\infty \!\!\! dp^+
%	\nfd(k^+{+}p^+) [1{-}\nfd(p^+)] 
	\frac{\nfd(k^+{+}p^+) [1{-}\nfd(p^+)] }{\nfd(k^+)}
\left[ \frac{-2k^-}{\mm}- \frac{(p^+)^2 + (p^+{+}k^+)^2}{k^+p^+ (p^+{+}k^+)}
	 \right].
	 \label{borndis}
\end{equation}
The kinematic edge occurs at $k^-=0$; but for $|k^- k^+| \ll \mm$ $p^+$
must be large, leading to exponentially small statistical factors.
Also note that the second term in square brackets is negative.
It actually dominates at large $p^+$, and
$\Pi^<_\mathrm{coll,Born,DIS}$ itself is negative for small values of
$|k^-|$.  This apparently unphysical result is because the negative
second term in \Eq{borndis} arises from the real part of a quark HTL
self-energy and vertex; the imaginary parts of the HTLs appear in the
$\twotwo$ processes, and the sum is always positive as it should be.

To understand the behavior at larger $K^2$ and in particular to make
contact with the calculations
\cite{Laine:2013vma,Ghisoiu:2014mha} which have been carried out there,
it is instructive to expand \Eq{fpborntot} in large $-K^2/\mm$.  The first step is to
estimate the importance of including the collision operator in
\Eq{transverseborn}.  It provides a width to the on-shell peak, and it
provides a small imaginary part off-peak.  Both effects are suppressed
by $\OO([\mm/(-K^2)]^2)$.  So at NLO in this expansion we can neglect
the collision operator, and start with \Eq{fpborntot}.
The $k^-/\mm$ term in \Eq{fpborntot} is larger than the
$\frac{(p^+)^2+(p^+{+}k^+)^2}{2k^+ p^+ (p^+ {+} k^+)}$ term by
one factor of $-K^2/\mm$, so the leading-order contribution is obtained
by integrating $k^-/\mm$ from $p^+=-k^+$ to $p^+=0$ (the small-$\mm$ limits
of $p^+_\mathrm{min,max}$), obtaining
\begin{equation}
\Pi^<_{\coll,\mathrm{free}}(K)
 = \frac{\Bb}{\mm}2 k^-T (1+2\nbe(k^+))
 \ln \left(\cosh\: \frac{k^+}{2T} \right),
\label{BornLOsemicoll}
\end{equation}
which we have labeled ``collinear free'' because it is precisely the Born term
that would arise in the free $g \to 0$ limit and in the collinear
limit $k^-/k^+\to 0$.

At the next order we must account for the narrower limits of the $p^+$
integration in this dominant term;
$p^+_\mathrm{max} \simeq -\frac{\mm}{2k^-}$ and
$p^+_\mathrm{min}= -k^+-p^+_\mathrm{max}$.  This cuts off each edge of
the integration range, reducing the result by
\begin{equation}
 \Pi^<_{\coll,\mathrm{Born},\mathrm{lim}}(K)= -\frac{\Bb}{\nfd(k^+)}
\times 2 \frac{\mm}{2k^-} \times
\nfd(k^+) [1{-} \nfd(0)] \frac{k^-}{\mm}
=-\frac{\Bb}{2}.
\label{borndelta}
\end{equation}
Here $\frac{\Bb}{\nfd(k^+)}$ is the prefactor in \Eq{fpborntot},
$2 \frac{\mm}{2k^-}$ is the amount of the integration region which
is removed, and the remaining expression is the integrand evaluated at
the edge ($p^+=0$ or $p^+=-k^+$).  We must also account for the
subdominant term in the integrand of \Eq{fpborntot}:
\begin{equation}
\label{bornthetadef}
\Pi^<_{\coll,\mathrm{Born},\theta}(K) =  \Bb
\int_{\frac{-k^+}{2}}^{\frac{-\mm}{2k^-}} dp^+
\frac{\nfd(k^+{+}p^+) [1-\nfd(p^+)]}{\nfd(k^+)}
\frac{(p^+)^2 + (p^+{+}k^+)^2}{k^+p^+ (p^+{+}k^+)},
\end{equation}
where the extrema arise from having exploited the symmetry around
$p^+=-k^+/2$ and in having expanded $p^+_\mathrm{max}$ as before.
At small $p^+$ the statistical functions become $\frac 12$ and
$\frac{(p^+)^2 {+} (p^+{+}k^+)^2}{k^+p^+(p^++k^+)}\simeq \frac{1}{p^+}$,
contributing a log to the integral.  Adding and subtracting this
limiting behavior, we find
\begin{eqnarray}
\label{borntheta}
\Pi^<_{\coll,\mathrm{Born},\theta}(K) &=&
 \Bb\left\{\frac{1}{2}\ln\left(\frac{\mm}{k^+k^-}\right)
\right.  \\  && \left. \nn \hspace{1.5em}
 - \int^{k^+/2}_{0}dp^+\left[\frac{\nfd(k^+{-}p^+) \nfd(p^+)} {\nfd(k^+)}
\frac{(p^+)^2 + (k^+{-}p^+)^2}{k^+p^+ (k^+{-}p^+)}
         -\frac{1}{2p^+}\right]\right\} .
\end{eqnarray}
The integration on the r.h.s, which can now be safely pushed to $p^+=0$, needs to be performed numerically. 
We note that the same integration was dealt with
in the context of the NLO real photon rate in \cite{Ghiglieri:2013gia} and an accurate fit of the result
as a function of $k^+/T$ was provided. Making up for slight differences in notation we have
\begin{equation}
	\int^{k^+/2}_{0}dp^+\left[\frac{\nfd(k^+{-}p^+)\nfd(p^+)}{\nfd(k^+)}\frac{(p^+)^2 + (k^+{-}p^+)^2}{k^+p^+ (k^+{-}p^+)}-\frac{1}{2p^+}\right]=
	\frac{1}{2}\left(C_\mathrm{pair}\left(\frac{k^+}{T}\right)-\ln\frac{k^+}{2T}\right),
\end{equation}
where the fit for $C_\mathrm{pair}$ can be read off from Eq.~(D.17) in \cite{Ghiglieri:2013gia}.
Our result agrees with the one in \cite{Ghisoiu:2014mha}.

Combining the pieces,
\begin{eqnarray}
\label{borntotexpanded}
\Pi^<_\mathrm{coll,Born}(K) & = & \Pi^<_\mathrm{coll,free}(K)
 + \Pi^<_{\coll,\mm}(K) + \OO([\mm/K^2]^2)\,,
\vphantom{\Big|} \nn \\
\Pi^<_{\coll,\mm}(K) & \equiv &
   \Pi^<_\mathrm{coll,Born,lim}(K)
 + \Pi^<_{\mathrm{coll,Born,}\theta}(K) \,.
\end{eqnarray}

Similarly, if we expand \Eq{borndis} in $|k^- k^+| \gg \mm$, we find (at
leading order)
\begin{equation}
	\Pi^<_\mathrm{coll,free,DIS}(K^2) =  -\frac{\Bb}{\mm}2k^-T
        (1+2\nbe(k^+))\ln\frac{2}{1+e^{-k^+/T}}.
	\label{freecolldis}
\end{equation}
Note that \Eq{fpborntot} is dominated by $p^+\sim k^+/2$, whereas
\Eq{borndis} is dominated by $p^+ \leq T$.  Therefore for $k^+\gg T$ the
collinear expansion works much better in the dilepton than in the DIS
regime, and \Eq{freecolldis} will have a narrower range of validity as
we go to larger $|k^-|$.

\section{Results}
\label{sec:results}
Let us start by collecting the leading order result, which, as we
argued, is given by the $\twotwo$ rate, unaffected by a nonzero $k^-$,
and the collinear rate. In detail
\begin{equation}
	\label{totallo}
	\Pi^<_{LO}(K)=\Pi^<_{\twotwo,LO}(K)+\Pi^<_{\coll}(K)=
	\Bb \left[ \ln \frac{T}{\justm}
	  + C_\mathrm{hard}\left( \frac{k^+}{T} \right) 
  + C_\coll\left( \frac{k^+}{T},\frac{k^-}{T},
    \frac{\mm}{\md^2} \right)\right],
\end{equation}
where $\Pi^<_{\twotwo,LO}(K)$ was introduced and $C_\mathrm{hard}$ was
given in \Eq{Chard} and $\Pi^<_{\coll}(K)$ (and correspondingly
$C_\coll$) is given by solving \Eq{fpprob2}.
We have made clear the dependence 
of collinear processes on $k^-$ and on $\mm/\md^2 = 3\cf/(4(\nc+\nf/2))$.

At NLO we must combine the soft and semi-collinear contributions, given by
\Eq{Csoft+sc}, with the collinear ones, obtained by perturbing 
$\Pi^<_{\coll}(K)$ with $\delta \mathcal{C}$ and $\delta \mm$, analogously to 
\Eq{df_NLO}. In detail
\begin{equation}
	\Pi^<_{NLO}(K)=\Pi^<_{LO}(K)+\delta\Pi^<(K),
	\label{defdeltapi}
\end{equation}
with the $\OO(g)$ correction $\delta\Pi^<(K)$ reading
\begin{equation}
	\label{totnlo}
	\delta\Pi^<(K)=\Bb\left[\frac{\delta\mm}{\mm}\ln\frac{\sqrt{2\md T}}{\justm}+\frac{\delta\mm}{\mm}C_\mathrm{soft+sc}\left(\frac{k^+}{T}\right)+
	\delta C_\coll\left( \frac{k^+}{T},\frac{k^-}{T},\frac{\mm}{\md^2} \right)\right],
\end{equation}
with $C_\mathrm{soft+sc}$ given in Eq.(6.3) of
Ref.~\cite{Ghiglieri:2013gia}.  These first two terms come from the
soft+semi-collinear contribution, \Eq{Csoft+sc}, and are the same
as for the real photon case.
$\delta \Ccoll$ arises from solving 
\Eq{df_NLO} modified by using \Eq{deltaE2} for
$\delta E$, plus an analogous contribution from $\delta g$. Finally,
we remark that $\mm/\md^2$ in Eq.~\eqref{totnlo} is to be intended
at LO, as given in the previous paragraph.
\subsection{Formal weak-coupling limit}

First consider the formal weak coupling limit, where all results have
been derived.  In this case $k^- \sim g^2 T$ can be neglected compared
to $k^+ \sim T$ in evaluating all statistical functions, and we need
not distinguish between $k$, $k^+$, and $k^0$.  Then we can evaluate
$\Ccoll$ and $\delta \Ccoll$ using \Eq{fpprob2} to
\Eq{deltaE2} and with \Eq{df_NLO} (and an analogous expression for
$g$).

%[[ FILL:  Fix the labeling in this figure, which is almost surely
%wrong!  Same with the figure of NLO corrections!!!]]
\begin{figure}
\includegraphics[width=0.45\textwidth]{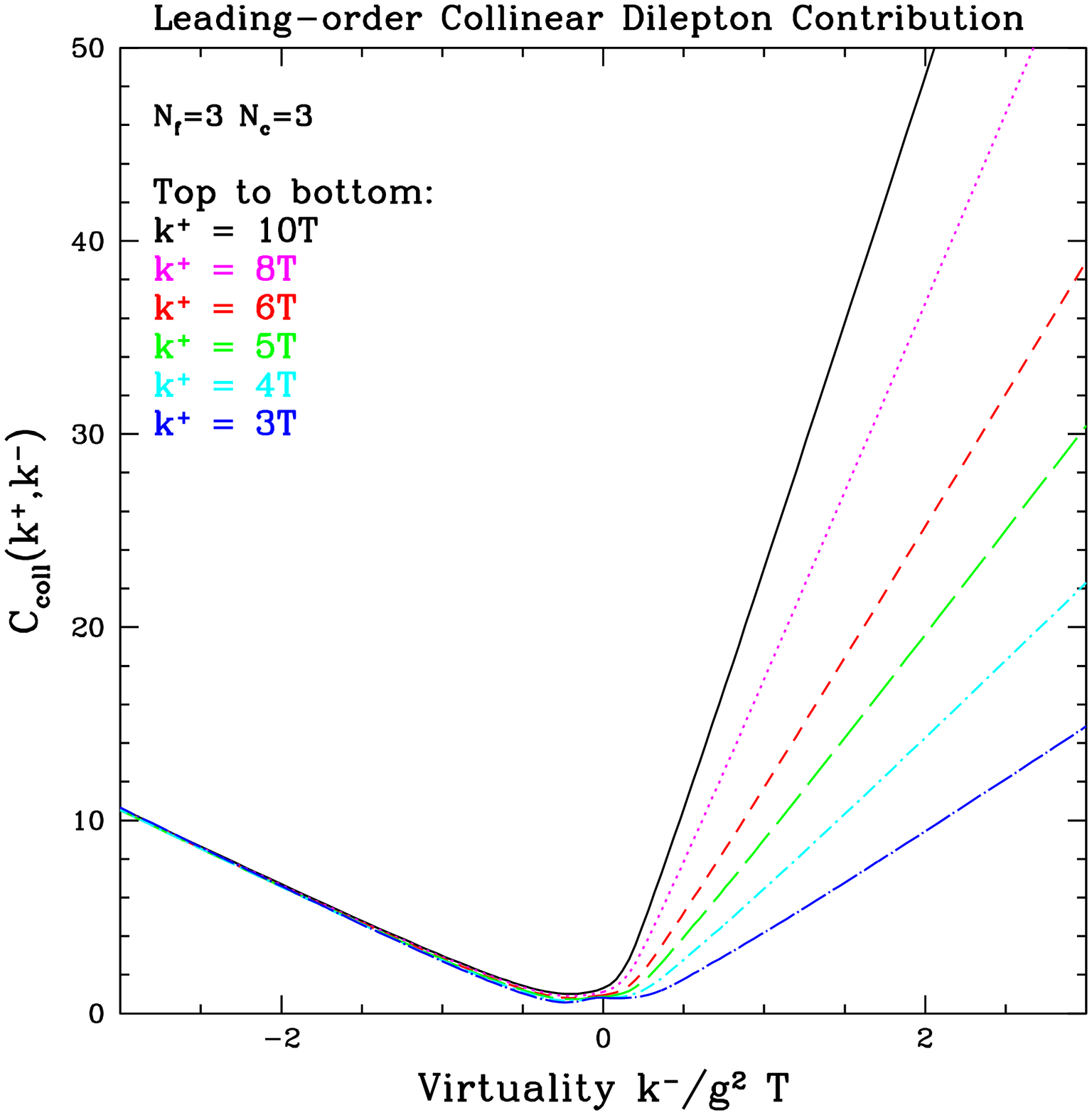}
\hfill
\includegraphics[width=0.45\textwidth]{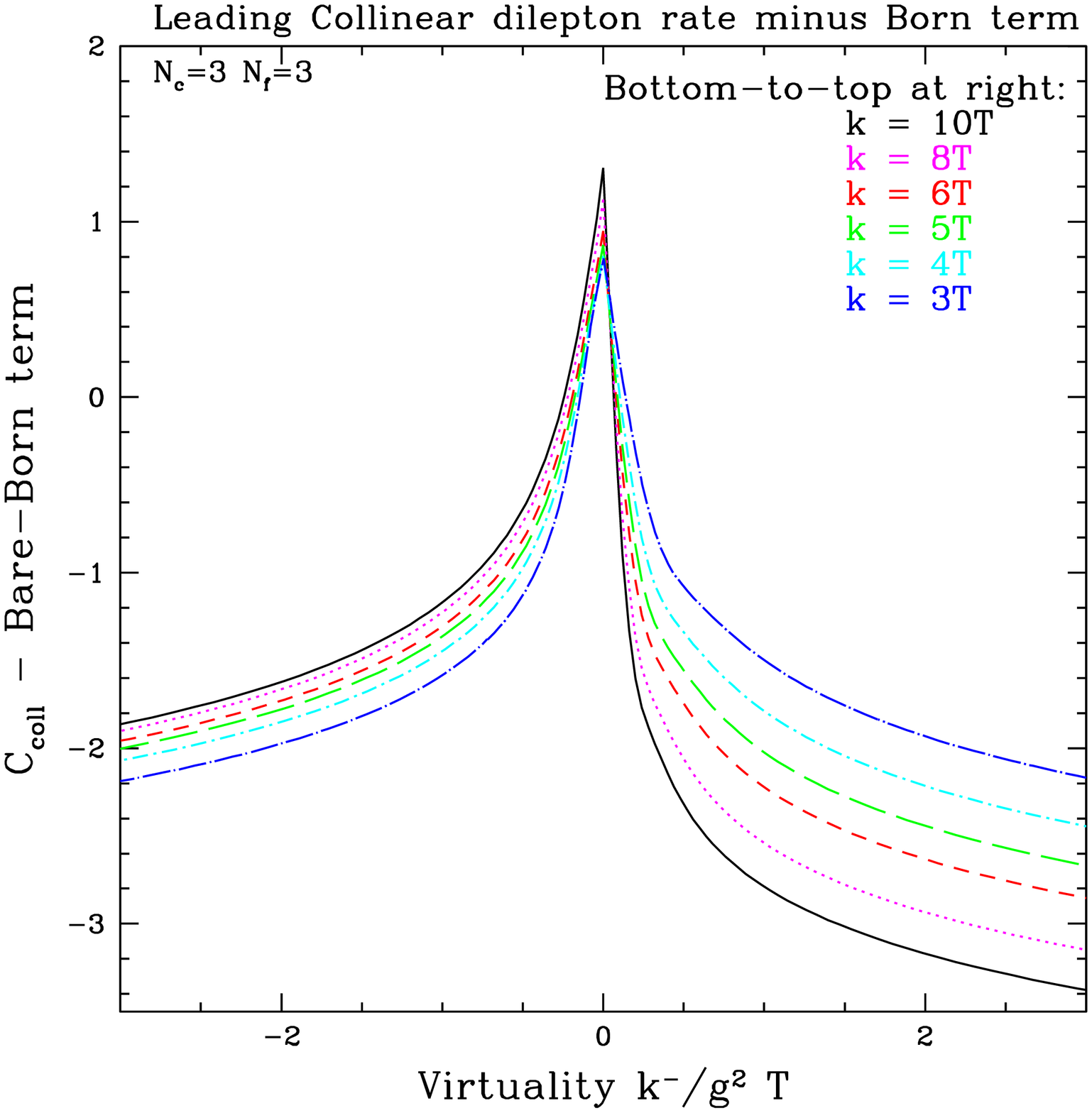}
\caption{\label{LOresultfig}
Contribution of collinear processes to dilepton production at leading
order, expressed as $\Ccoll$.  That is, $\Pi^<$ is given by
$\Bb$ times the indicated value.  In the left plot we display the full
value; on the right side we subtract the (free) Born contribution,
\Eq{BornLOsemicoll} and \Eq{freecolldis}.  The cusp in the right plot is because the
Born term has a cusp (see \Fig{fig_bare}) which is subtracted from a
smooth function.}
\end{figure}

\begin{figure}
\begin{center}
\includegraphics[width=0.45\textwidth]{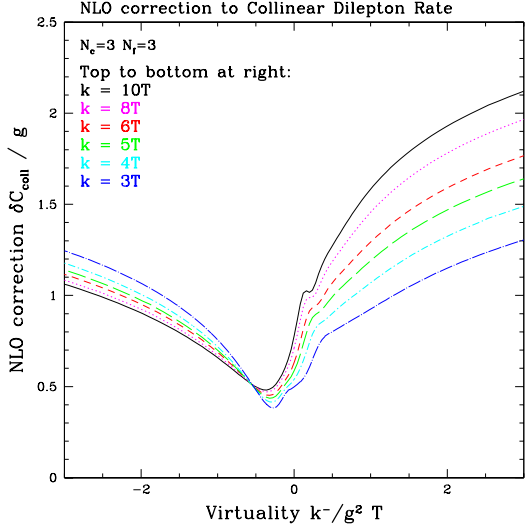}
\end{center}
\caption{\label{NLOresultfig}
Subleading correction to the collinear contribution, expressed as
$(1/g) \delta \Ccoll$.  In an NLO computation, $g$ times this
result should be added to $\Ccoll$.}
\end{figure}

We present results for the specific case $\nf = 3$ in
\Fig{LOresultfig} and \Fig{NLOresultfig}, which show
$\Ccoll$ and $(1/g) \delta \Ccoll$ respectively at a
few values of $k^+$.  We also tabulate the results in Appendix
\ref{AppTab}, and provide a code for evaluating $\Ccoll$ and $\delta
\Ccoll$ in the extra included files.  \Fig{LOresultfig} shows that
the result rapidly approaches the Born value as $k^-/g^2 T$ becomes
large. Therefore we have also presented the difference between the LO
result and the Born value.  The subtraction introduces a cusp at
$k^-=0$, since the Born value possesses such a cusp. As
\Fig{NLOresultfig} shows, the NLO correction always increases the
rate, but it is a complicated function of $k^-/g^2 T$.

\subsection{Treating the coupling as finite}

Phenomenologically, we are interested in obtaining results for
specific finite values of $g$ -- especially we want to push towards
values of physical interest in heavy ion collisions, even though the
coupling expansion will break down in this limit.  When treating $k^-$
as formally infinitesimal, we did not need to distinguish between $k$,
$k^0$, and $k^+$ at NLO.  However we must make some sensible choice in
which one to use when we treat $k^-$ as a finite number times $T$.
Our choice is the following.  Consider a plot of $\Pi^<$ for a fixed $k$
as a function of $k^0$.  We treat $\Bb = \Bb(k^0)$ (so the main
statistical function controlling the overall size of our results is in
terms of $k^0$ as it should be).  Then we treat $k^+$ as equal to $k$
where they appear inside the integral in \Eq{fpprob2} and
in \Eq{qspace2}, \Eq{migdallong}, and \Eq{deltaE2}; and $k^-$
is determined as expected, $k^- = k^0 - k$.

In addition, once we start considering $k^- \sim T$, the collinear
approximations we have made stop being as accurate.  In this regime it
makes more sense to perform an expansion which treats $k^-$ and $k^+$
on equal footing.  Of course, this has already been accomplished
by Laine \cite{Laine:2013vma}.  So it would be valuable to improve our
result in this way.  As we have shown in \Eq{BornLOsemicoll},
the large-$k^-$ limit of the collinear part is enhanced by
$k^+k^-/\mm$ over the remainder of the LO result, giving rise to the
collinear free (Born) term.  We can subtract this term and add the free Born
term, computed taking finite angles into full account.  The
finite-angle Born term is well known and reads \cite{McLerran:1984ay}
\begin{equation}
 \Pi^<_{\mathrm{free}}(K) = \frac{\Bb} {\mm\nfd(k^0)}
  \frac{k_0^2-k^2}{k}T\, \nbe(k^0)\,
 \ln \left(\frac{\cosh\left(\frac{k^+}{2T}\right)}
        {\cosh\left(\frac{k^-}{4T}\right)}\right),
\label{Bornfull}
\end{equation}
which indeed reduces to \Eq{BornLOsemicoll} at first order in
$k^-/k^+$ and is exactly the leading-order result in the $K^2\sim T^2$
calculation of \cite{Laine:2013vma}. Similarly, in the DIS region it
reads
\begin{equation}
 \Pi^<_{\mathrm{free,DIS}}(K) = -\frac{\Bb} {\mm\nfd(k^0)}
  \frac{k_0^2-k^2}{k}T\, \nbe(k^0)\, 
  \ln \left(\frac{1+e^{k^-/(2T)}}
  {1+e^{-k^+/T}}\right),
\label{disfull}
\end{equation}
which again reduces to \Eq{freecolldis} at first order in $k^-/k^+$.

Therefore, our first step in making our results reliable at larger
$k^-$ is to subtract \Eq{BornLOsemicoll} and replace it
with \Eq{Bornfull}. That leads to\footnote{For $k^-<0$, i.e. in
	the DIS region,
	the subtractions and additions are understood to be
	of the appropriate DIS terms, that is $\Pi^<_\mathrm{free,DIS}-
	\Pi^<_\mathrm{coll, free,DIS}$ in
place of $\Pi^<_\mathrm{free}-\Pi^<_\mathrm{coll,free}$.}
\begin{equation}
	\Pi^<_{LO,\,\mathrm{large}}(K)=\Pi^<_{LO}(K)
  - \Pi^<_{\coll,\mathrm{free}}(K)+ \Pi^<_{\mathrm{free}}(K),
\label{totallosubtr}
\end{equation}
and leaves the NLO correction $\delta \Pi^<$ unmodified, so that
\begin{equation}
\Pi^<_{NLO,\,\mathrm{large}}(K)=\Pi^<_{LO\,\mathrm{large}}(K)+\delta\Pi^<(K).
\label{defdeltapilarge}
\end{equation}
Such a prescription will thus ensure that the leading, large-$K^2$ behavior of 
our LO and NLO calculations agrees with the leading order in \cite{Laine:2013vma}.
The next-to-leading order corrections to that calculation arise from the diagrams
in \Fig{fig_diagrams}, evaluated without resummations for all possible cuts, thus
including both virtual corrections to the Born term and real corrections in the form
of $\twotwo$ processes, properly taking into account the no-longer negligible virtuality.
A procedure to patch that calculation with an LPM-resummed one for
small $k^-$ was recently introduced by Ghisoiu and Laine in
\cite{Ghisoiu:2014mha}. From the point of view of our calculation, it
requires, on top of the previous substitution of the free term, the subtraction of the
remainder of the large-$K^2$ behavior of the leading order, \textsl{i.e.},
$\Pi^<_{\coll,\mathrm{Born},\mm}(K)$ and $\Pi^<_{\twotwo,LO}$, and its
replacement with the NLO correction for large $K^2$ obtained by Laine
in \cite{Laine:2013vma}. In detail,
\begin{eqnarray}
	\Pi^<_{LO,\,\mathrm{GL}}(K)&=&\Pi^<_{LO,\,\mathrm{large}}(K)-\Pi^<_{\coll,\mathrm{Born},\mm}(K)-\Pi^<_{\twotwo,LO}(K)+\Pi^<_{NLO,\mathrm{Laine}}(K)\nn\\
	&=&\Pi^<_{\coll}(K)- \Pi^<_{\coll,\mathrm{free}}(K)-\Pi^<_{\coll,\mathrm{Born},\mm}(K)+\Pi^<_{\mathrm{free}}(K)+\Pi^<_{NLO,\mathrm{Laine}}(K),\nn\\
	&&  \label{totallosubtrmikko}
\end{eqnarray}
where $\Pi^<_{NLO,\mathrm{Laine}}$ is the $\OO(g^2)$ correction to $\Pi^<_{\mathrm{free}}(K)$
computed by Laine. As shown there, it diverges as $\ln(T^2/-K^2)$ at small $K^2$. Similarly,   
$\Pi^<_{\coll,\mathrm{Born},\mm}(K)$ contains a $\ln(\mm/-K^2)$, which is correct at large
$k^-$ but divergent and unphysical at small $k^-$. The subtraction leaves a $\ln(T^2/\mm)$,
which is exactly the logarithmic term we encounter at LO.

\begin{figure}[ht]
\begin{center}
\includegraphics[width=12cm]{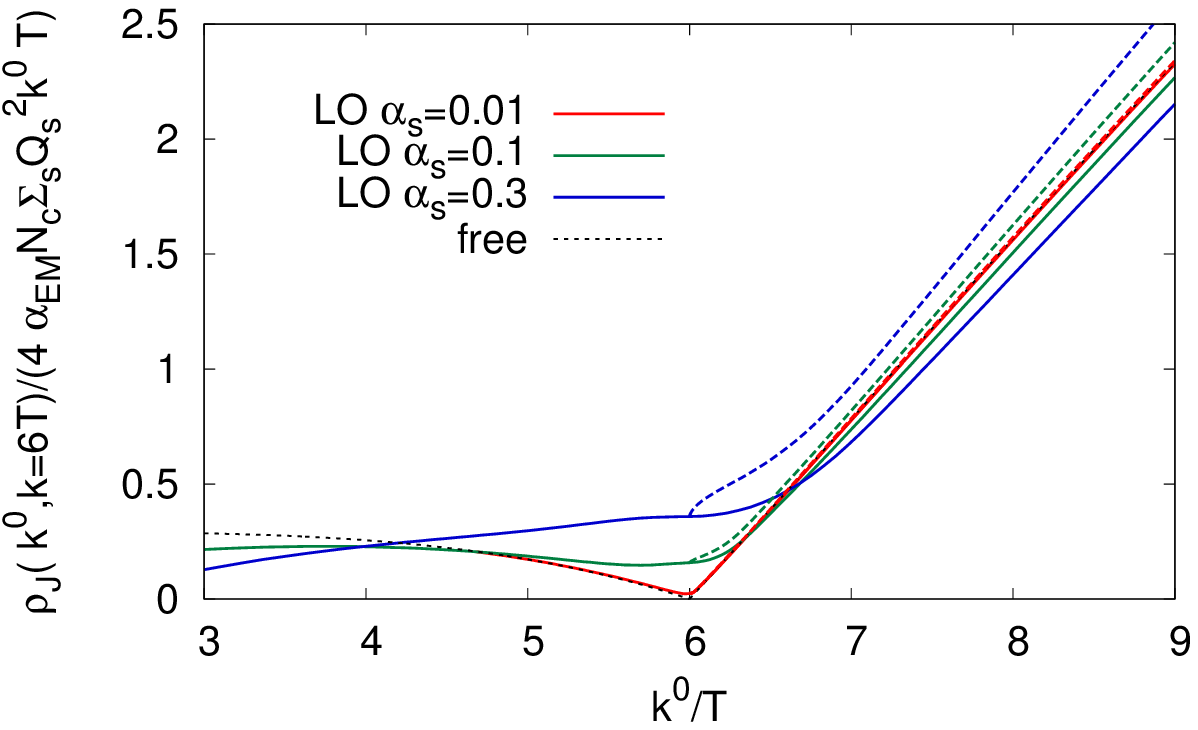}	
\end{center}
\vspace{-1cm}
\caption{Plot of $\rho/(k^0/T) = (\Pi^< /\Bb)(\nfd(k^0)/\nbe(k^0))$
for 3 light flavors at fixed $k=6T$ as a function of $k^0$:  for
$k^0<6T$ this determines the deep
inelastic scattering rate, for $k^0>6T$ it determines the dilepton rate,
$k^0=6T$ sets the real photon rate.  Solid lines use
$\Pi_{LO,\mathrm{large}}$ (\Eq{totallosubtr}), while the dotted lines use
$\Pi_{LO,\mathrm{GL}}$ (\Eq{totallosubtrmikko},\cite{Ghisoiu:2014mha}),
which is only available for $k^0>k$.}
\label{fig_lo_6}	
\end{figure}

\begin{figure}[ht]
\begin{center}
\includegraphics[width=12cm]{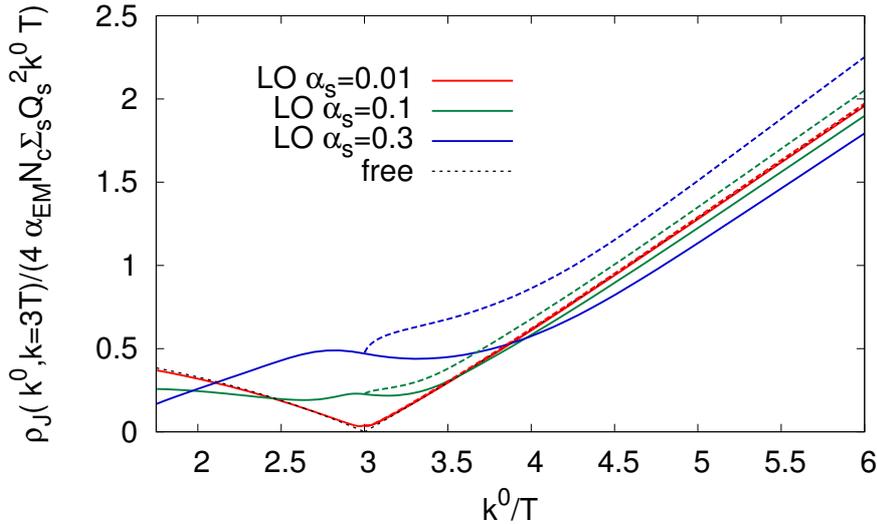}	
\end{center}
\vspace{-1cm}
\caption{The same as \Fig{fig_lo_6}, but for the case $k=3T$.}
\label{fig_lo_3}	
\end{figure}

Figures~\ref{fig_lo_6} and \ref{fig_lo_3} show a comparison of the two leading-order
prescriptions \eqref{totallosubtr} and \eqref{totallosubtrmikko} for two
different values of $k/T$ for the case of 3 light flavors and at a few
gauge couplings. For practical reasons [to eliminate a trivial
overall $e^{-k^0/T}$ behavior], we plot the 
spectral function $\rho_J$ rather than $\Pi^<$ (we recall that, at equilibrium,
$\Pi^<(K)=\nbe(k^0)\rho_J(K)$). In both figures, the continuous lines represent
the results obtained from $\Pi^<_{LO,\mathrm{large}}$ at three different
values for the coupling, whereas the dashed lines correspond
to $\Pi^<_{LO,\mathrm{GL}}$, which is unfortunately only available for
$k^0>k$.

Interestingly, the two prescriptions agree at $k^0=k$ but seem to deviate significantly already at very 
small $k^-$. We believe that at small $k^-$ the discrepancy can be attributed in part to
our arbitrary choice regarding the $k^0$, $k^+$, $k$ ambiguity, in part to the subleading
dependence of the $\twotwo$ processes on $k^-$, which we omit but Laine includes\footnote{At finite $k^-$ one also has $3\leftrightarrow 1$ processes, which are also included in Laine's calculation.}, and in
part to a potential unreliability of the Ghisoiu-Laine procedure at
$-K^2 \ll 4\mm$ (see footnote~5
in \cite{Ghisoiu:2014mha}). We furthermore remark that in App.~\ref{app_gt2}
we show by an explicit calculation that, for $k^-\sim gT$, 
$\Pi^<_{LO,\mathrm{large}}$ is correct at $\OO(e^2g^2T^2)$, that is, our
calculation is fully NLO in this regime as well as for $k^- \sim g^2 T$.

On the other hand, a significant deviation at larger values of $k^-/T$ is to be expected
and there the reliable curve is certainly the dashed one, which for $k^+,k^-\gg T$ is dominated
by the (finite-angle) Born term and its finite vacuum corrections, which are completely absent from
our calculation. Conversely, our calculation asymptotes in this region to the Born term
supplemented by a $g^2 T^2$  and a $g^2 T^2\ln(kk^-/T^2)$
correction. Note that, at least for very large $k^- \gg T$, we know based on
Operator Product Expansion
techniques \cite{CaronHuot:2009ns}
that $g^2 T^2$ terms, with or without accompanying logarithms, will be absent.

\begin{figure}[ht]
\begin{center}
\includegraphics[width=12cm]{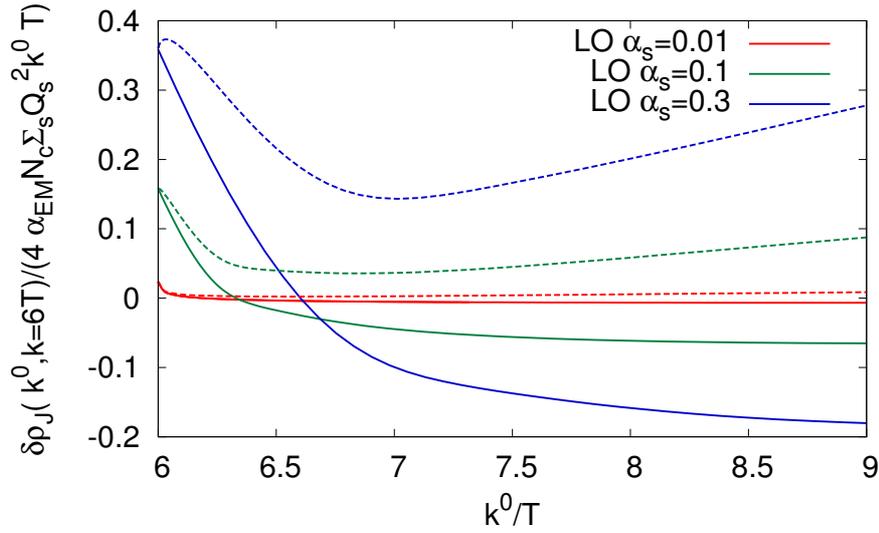}	
\end{center}
\vspace{-1cm}
\caption{The same as \Fig{fig_lo_6}, but with the dominant Born
  term subtracted.}
\label{fig_lo_6_noborn}	
\end{figure}

In order to better highlight the differences between the two prescriptions,
in \Fig{fig_lo_6_noborn} we subtract off the Born term
$\Pi_{\mathrm{free}}^<$ from each curve in \Fig{fig_lo_6}.
One can then more easily see the deviations at small $k^-$ at intermediate
and large coupling, as well as the emergence of the vacuum correction to the
Born term in the dashed lines that give rise to the approximate linear behavior at larger
$k^0$.

\begin{figure}[ht]
\begin{center}
\includegraphics[width=12cm]{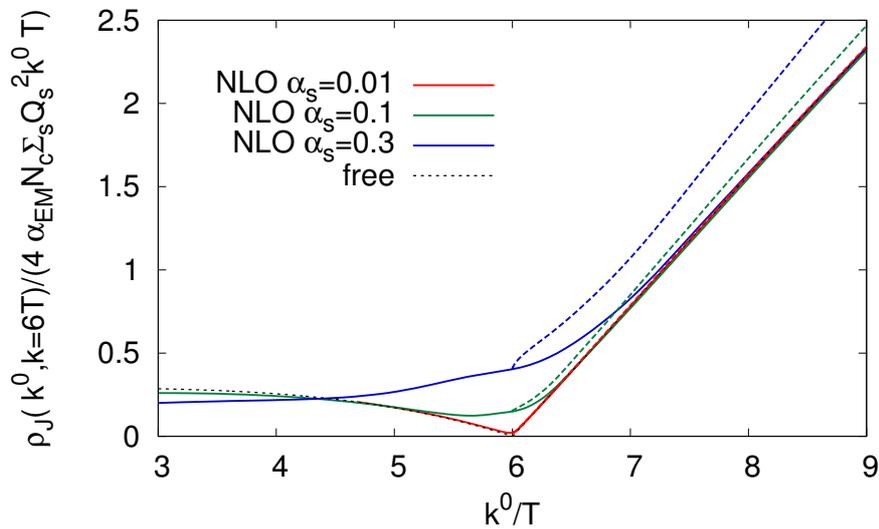}	
\end{center}
\vspace{-1cm}
\caption{The same plot as \Fig{fig_lo_6}, but with the inclusion of the
  small-virtuality NLO corrections which are the focus of this paper.
  The solid curves are \Eq{defdeltapilarge} and the dotted curves are
  \Eq{totalnlosubtrmikko}.}
\label{fig_nlo_6}	
\end{figure}

\begin{figure}[ht]
\begin{center}
\includegraphics[width=12cm]{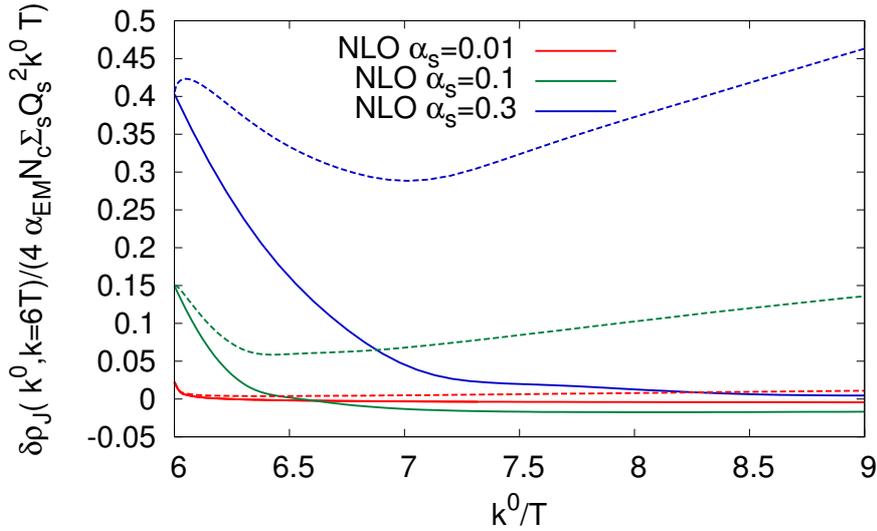}	
\end{center}
\vspace{-1cm}
\caption{The same as \Fig{fig_nlo_6} but subtracting the dominant free
  Born term as in \Fig{fig_lo_6_noborn}.}
\label{fig_nlo_6_noborn}	
\end{figure}

Of course, the most accurate estimate of the small virtuality region we
can make is the one where we also include the NLO corrections we have found:
\begin{equation}
	\Pi^<_{NLO,\mathrm{GL}}(K)=\Pi^<_{LO,\mathrm{GL}}(K)+\delta\Pi^<(K).
	\label{totalnlosubtrmikko}
\end{equation}
Note that none of the effects which give rise to $\delta \Pi^<(K)$
involve diagrams which are included in Laine's calculation
\cite{Laine:2013vma}, and no NLO contribution grows as a power%
\footnote{%
    At large $k^-$ the largest NLO term is
    the $\delta\mm$ contribution to Eq.~\eqref{borntheta}
    which grows logarithmically with $k^-$.}
at large $k^-/g^2 T$, so no new subtraction is called for here.
Figures~\ref{fig_nlo_6} and \ref{fig_nlo_6_noborn} are then the NLO counterparts of
Figs.~\ref{fig_lo_6} and \ref{fig_lo_6_noborn}, with $\delta\Pi^<$ added to
both full and dashed lines.  These represent our best estimate of the
spectral function relevant for dilepton production in the small
virtuality region.

\begin{figure}[ht]
\begin{center}
\includegraphics[width=12cm]{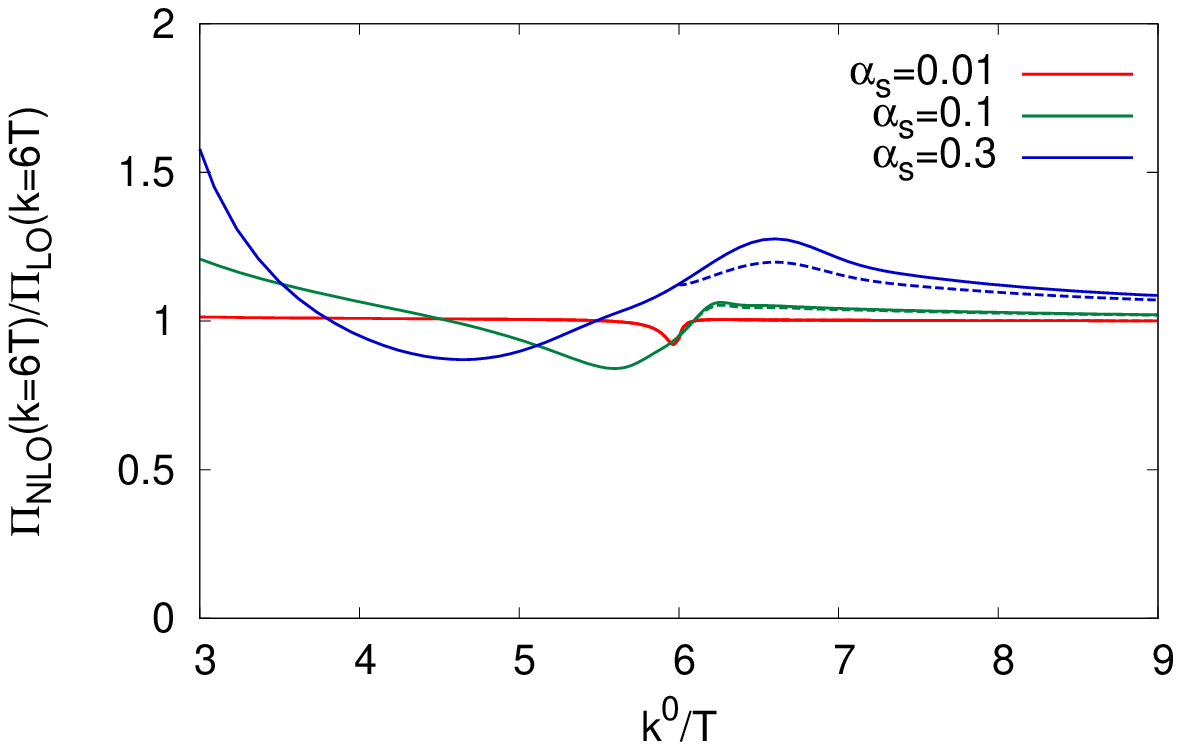}	
\end{center}
\vspace{-1cm}
\caption{The NLO/LO ratio for $k=6T$. Solid lines are the ratio of
\Eq{defdeltapilarge} to \Eq{totallosubtr}, while dashed lines are the
ratio of \Eq{totalnlosubtrmikko} to \Eq{totallosubtrmikko}.}
\label{fig_ratio_6}	
\end{figure}

In order to highlight the effect the NLO corrections have on the LO rate, in 
\Fig{fig_ratio_6} we plot the ratio $\Pi^<_{NLO}/\Pi^<_{LO}$ in the two 
prescriptions, with the same convention for distinguishing them.
At the smallest coupling the NLO effect is negligible except for a 
dip of a few percent across the light cone, consistently with what is observed
in the real photon case. At larger couplings the dip disappears, to be replaced
by a maximum at positive $k^-$ and a minimum on the opposite side. The large corrections observed
at negative $k^-$ are not to be trusted; as discussed below
\Eq{freecolldis}, the collinear approximations in our treatment break
down for $k^- < -T$.

\section{Conclusions}
\label{sec:conclusions}
We have computed to next-to-leading order the production
rate for dileptons with a small virtuality, parametrically of 
order $g^2T^2$. This required a careful merging of the 
 NLO result for real photons and of the LO result for small-mass
 dileptons. To this end, we have reviewed these calculations
 in Sec.~\ref{sec:review}. We showed how the LO photon rate 
 arises from two distinct processes, elastic scattering and 
 collinear, inelastic splitting, corresponding to separate kinematical regions.
 The LO dilepton rate at small $K^2$ requires only a modification
 of the latter processes to account for the non-vanishing virtuality.
When computing the photon rate to NLO, the two processes and the 
corresponding kinematical regions start to blur and merge, requiring careful
analyses which we reviewed. In particular, an intermediate process/region arises, the
semi-collinear one.

In Sec.~\ref{sec:NLOdilepton} we have shown that, similarly to LO, the extension
of the NLO photon rate to finite virtuality requires only a modification of the 
purely collinear part of the NLO photon rate. The effect of the finite virtuality
on the other processes is negligible to this order. The collinear rate to leading- 
and next-to-leading order is obtained by solving a set of differential, Schr\"odinger-like
equations that resum the effects of the multiple collisions of the quarks that emit/annihilate
into the virtual photon with the lightlike medium constituents.
We have shown in detail
how the equations governing the photon rate are modified and we have introduced
in App.~\ref{app_hankel} a new technique for solving them. We also provide code and 
tables of the results in App.~\ref{AppTab}. We have furthermore shown that the zeroth
order in the resummation of collisions, \textsl{i.e.}, no collisions, is (the collinear limit of)
the Born term, that is the direct annihilation of a quark and an antiquark into the virtual
photon. Due to the presence of thermal masses for the quarks, such a term only exists for
$-K^2>4\mm$, and it becomes dominant when $-K^2\gg 4\mm$.

Numerical results from the solution of these equations, together with the other terms
that are taken, unmodified, from the photon calculations, are presented in
Sec.~\ref{sec:results}. We remark that, while the dilepton rate is defined
only for positive $k^-$, the related electromagnetic Wightman function
$\Pi^<$ is defined for any $k^-$ and our calculation is valid for $k^-\sim
g^2T$, irrespective of its sign. We are then able to analyze the behavior
of this function across the light cone, which is of importance also
for the ongoing efforts at reconstructing this function from lattice measurements
of Euclidean correlators.

We find that at small couplings $\Pi^<$ resembles the free one shown in 
\Fig{fig_bare}, with the cusp smoothed by the small ($\OO(g^2T^2)$)
finite photon rate. At larger couplings the transition from the spacelike
to the timelike region is smoother, resembling more the results obtained
at strong coupling in SYM. These findings are highlighted by Figs.~\ref{fig_lo_6} to
\ref{fig_nlo_6_noborn}. These figures also show the two different prescriptions
we have adopted to stretch the validity of our calculation to larger values of $k^-$. The first
entails the replacement of the collinear limit of the Born term
with its full counterpart, whereas the second, originally due to Ghisoiu and Laine, patches
Laine's calculation at large $k^-$ with the LPM-resummed one at small $k^-$.
We observe deviations between the two prescriptions. At large $k^-$ the latter one is to be 
trusted, whereas at smaller values we believe the first one to be more accurate.
The NLO corrections we have found generally increase the dilepton rate,
by up to 30-40\% for $\alphas = 0.3$ and small virtuality, but they make
little difference at larger virtuality.  
(This is in contrast to the large-virtuality NLO correction found by
Laine, which is more important at large virtuality but which is subsumed
into our LO calculation at small virtuality.)

The tabulated form and the code we provide can be used to readily incorporate
our results into phenomenological analyses taking into account the
hydrodynamical evolution of the near-equilibrium plasma produced in
heavy-ion collisions.

Finally, we should address two of the questions which motivated this
study.  First, experimentalists have sometimes interpreted the dilepton
rate at small mass but large energy as a valid substitute for the real
photon production rate.  For instance, in Ref.~\cite{EXPdilepton}, the
PHENIX collaboration considers di-electrons with $K^2 <
(300\:\mathrm{MeV})^2$ and $k \in [1,5]\:\mathrm{GeV}$.  Assuming for
the sake of argument a temperature of 250 MeV, this corresponds to
$k \in [4,20] T$, and $k^- = K^2/(2k^+) < 0.12\,T$ for $k^+=6T$.  So for
an estimate of whether $\Pi^<$ is the same for a real photon and a
dilepton pair in this virtuality range, we should compare the value of
the top solid (blue) curve in \Fig{fig_nlo_6} at $k^0/T=6$ and $k^0/T=6.12$.  The
solid blue curve shows a change of 4\%; the dashed curve (which is
probably the less reliable one in this range!) changes 25\%.  Even using
the unrealistically small $\alphas=0.1$ (green curves), the change is
20\% or 40\%.  Therefore we find that the difference in electromagnetic
response between photons and such small mass dileptons is negligible,
within our perturbative computational framework.  This supports the
interpretation of such low-mass dileptons as a legitimate replacement
for a real photon measurement.

The other question we wanted to address is whether to expect sharp
features in the spectral function, which could upset an analytical
continuation of lattice data.  \Fig{fig_nlo_6} suggests that the
spectral function ``turns on'' over a range of 1--2 $T$ width, and shows
no true ``notch'' feature.  This is much smoother than we might have
worried by looking at the free case, \Fig{fig_bare}.  Whether it is
smooth enough is a question we will leave to the experts in the field.
But we close with the remark that it may be easier to determine this
with a better understanding of the small virtuality behavior of
Laine's un-resummed NLO calculation \cite{Laine:2013vma} and an
extension of this calculation to $k^- < 0$ (the DIS regime).

\section*{Acknowledgments}

We would like to thank Aleksi Kurkela and Derek Teaney for
collaboration in the early stages of this work. We also thank
Mikko Laine for useful conversations.  This work was supported
in part by the Institute for Particle Physics (Canada), the Natural
Sciences and Engineering Research Council (NSERC) of Canada and
the Swiss National Science Foundation
(SNF) under grant 200020\_155935.  We
would also like to acknowledge the Mainz Institute for
Theoretical Physics (MITP) for enabling us to complete a portion of this work.

\appendix

\section{Numerical method}
\label{app_hankel}

Here we present an efficient approach to evaluating
\Eq{qspace2}, \Eq{migdallong} and in particular of evaluating
the required integrals in the second line of \Eq{fpprob2}.  Our starting
point is the observation of Aurenche \textsl{et al} \cite{AGMZ} that the
equations are most easily solved by Fourier transforming from
$\pp$ to $\b$ (impact parameter space).  The Fourier-transformed
versions of \Eq{qspace2} and \eqref{migdallong} read
\begin{eqnarray}
\label{bspace}
-2i \nabla \delta^2(\b)& =& \frac{i k^+}{2p^+(k^++p^+)} 
\Big( \mm +\frac{2p^+(k^++p^+)}{k^+}k^- - \nabla_b^2 \Big) \f(\b) + \cc(b) \f(\b) \,,\\
\label{migdallongIP}
2	\delta^2(\b)&=& \frac{i k^+}{2p^+(k^++p^+)} 
\Big( \mm +\frac{2p^+(k^++p^+)}{k^+}k^- - \nabla_b^2 \Big) g(\b) + \cc(b) g(\b) \,,
\end{eqnarray}
with $\cc(b)=\frac{g^2 \crr T}{2\pi}( K_0(b\md)+\gamma_{\sss{E}}+\ln(b\md/2))$.
We can make \Eq{bspace} a scalar equation by defining $\f(\b)=\b f(b)$,
\begin{eqnarray}
\label{bspacescalarized}
&&\frac{i k^+}{2p^+(k^++p^+)} 
\Big( \mm +\frac{2p^+(k^++p^+)}{k^+}k^- - \partial_b^2-3\frac{\partial_b}{b} \Big) 
f(b) + \cc(b) f(b)=0 \,,\\
\label{bspacescalarizedlong}
&&\frac{i k^+}{2p^+(k^++p^+)} 
\Big( \mm +\frac{2p^+(k^++p^+)}{k^+}k^- - \partial_b^2-\frac{\partial_b}{b} \Big) g(b) + \cc(b) g(b)=0 \,,
\end{eqnarray}
and implement the LHS of \Eq{bspace}, \Eq{migdallongIP} as
boundary conditions at $b\to 0$:
\begin{eqnarray}
\label{boundarytrans}
f(b\to 0) &=& \frac{2 p^+ (k^+{+}p^+)}{\pi k^+ b^2} +\OO(b^0)\,,\\
\label{boundarylong}
g(b\to 0)&=&\frac{2 i p^+(p^+{+}k^+)}{\pi k^+}\ln(b\md)+\OO(b^0)\,.
\end{eqnarray}
The quantities one has to solve for, appearing on the second line of
\Eq{fpprob2}, are 
\begin{eqnarray}
\label{wanttrans}
	\int \frac{d^2\pp}{(2\pi)^2} {\mathrm Re}\, \bpp \cdot \f(\bpp)
	 &=& {\mathrm Im}(  \nabla_{b} \cdot \f(b))
=2\,{\mathrm Im}\, f(0)\,,\\
\int\frac{d^2\pp}{(2\pi)^2}\mathrm{Re}\,g(\bpp)&=&\mathrm{Re}\,g(0)\,.
\label{wantlong}
\end{eqnarray}
A further boundary condition is given by the requirement that 
$f$ and $g$ decay to zero as $b\to\infty$.  Our strategy is to start
at large $b$ and use a standard ODE solver to evolve
\Eq{bspacescalarized} inwards, multiplicatively adjusting the solution
at $b\rightarrow 0$ so as to obey \Eq{boundarytrans}.

Let us define $m_\mathrm{eff}^2$ as
\begin{equation}
	\label{defmeff}
	m_\mathrm{eff}^2\equiv\mm +\frac{2p^+(k^++p^+)}{k^+}k^-\,.
\end{equation}
As shown in \cite{AGMZ}, generic solutions of $f$ and $g$ at
large values of $b$ are given by a superposition of
$\exp(\pm\sqrt{m_\mathrm{eff}^2}b)$, with slight perturbations caused by
$\cc(b\to\infty)\sim \log(b\md)$. As long as $m_\mathrm{eff}^2>0$, the
exponents are real and it suffices to initialize the numerical solver at
an appropriately large value of $b$ to pick the shrinking solution. On
the other hand, when the mass is negative, \textsl{i.e.}, where a Born
term is present (see \Eq{fpborntot}), the behavior at
large $b$ is mostly oscillatory, with a small damping introduced by the
presence of $\cc(b)$. This can cause stiffness in the numerical
solution, making numerical precision challenging.  This is especially so
for $-K^2 \gg 4\mm$.  It can also be numerically delicate to treat the
$f(b)$ equations near $b=0$, where \Eq{boundarytrans} shows the solution
blows up, but \Eq{wanttrans} shows we need to extract a finite piece.

We handle this problem by solving the ODE via a technique which takes
into account the dominant oscillating behavior analytically and solves
only for the interesting corrections due to $\cc(b),\delta \mm$.  We
start by rescaling $f(b),g(b)$ to bring \Eq{bspacescalarized},
\Eq{bspacescalarizedlong} into a standard form,
\begin{equation}
\left(  \partial^2_b + (d/b) \partial_b + A + i K(b)  \right) \tilde f(b) = 0.
\label{defgeneric}
\end{equation}
Specifically,
\begin{equation}
	\tilde f(b) = \frac{k^+}{2p^+(p^++k^+)}f(b),\qquad
	A = -m_\mathrm{eff}^2\,,\qquad
	K(b) = \frac{2p^+(k^++p^+)}{k^+}\cc(b),
	\label{defa}
\end{equation}
and similarly for $\tilde{g}$; $d=3$ for $f$ and $d=1$ for $g$.

The independent solutions to the $K(b)=0$ equation are known
analytically.  Call them%
\footnote{
  The solutions are Bessel functions up to $b$-rescaling, fractional
  powers of $b$, and possible normalization conventions.}
$J(b)$ and $Y(b)$, obeying
\begin{equation}
\left[ \partial_b^2 + \frac{d}{b} \partial_b + A \right]
 J(b) = 0\,,
\label{defJ}
\end{equation}
and likewise for $Y$, with Wronskian
\begin{equation}
\label{defwronski}
W(b) = Y(b) J'(b) - J(b) Y'(b) = c b^{-d}\,,
\end{equation}
with $c$ a constant which depends on the normalization convention for
$J,Y$.  For $A>0$ $J,Y$ are expressed in terms of Bessel and Neumann
functions, while for $A<0$ they involve growing and shrinking modified
Bessel functions $I,K$; but we will write $J,Y$ in the following for
notational simplicity.  In either case we choose $Y$ to be the divergent
and $J$ to be the finite solution at the origin.  The most general
solution to the $K(b)=0$ equation is $\tilde f(b) = c_1 J(b) + c_2
Y(b)$.  With this in mind, we parametrize the actual solution as
\begin{equation}
\tilde f(b) = c_1(b) J(b) + c_2(b) Y(b)\,.
\end{equation}
This form is underdetermined, as we have doubled the number of
freedoms.  We eliminate the extra freedom by imposing the condition
\begin{equation}
\label{extracondition}
J(b) c_1'(b) + Y(b) c_2'(b) = 0\,.
\end{equation}
Therefore
\begin{eqnarray}
\partial_b \tilde f(b) &=& c_1(b) J'(b) + c_2(b) Y'(b) \,, \nn \\
\partial_b^2 \tilde f(b) & = & c_1'(b) J'(b) + c_2'(b) Y'(b) 
       + c_1(b) J''(b) + c_2(b) Y''(b) \,,
\end{eqnarray}
which together with \Eq{defJ} turns \Eq{defgeneric} into the relation
\begin{equation}
0
 = i K(b) ( c_1(b) J(b) + c_2(b) Y(b) ) + c_1'(b) J'(b) + c_2'(b) Y'(b)\,.
\end{equation}
Multiplying by either $Y(b)$ or $J(b)$ and using \Eq{extracondition} and
\Eq{defwronski} we obtain coupled first-order equations for $c_{1,2}$:
\begin{eqnarray}
\nn c_1'(b) & = & -i \frac{Y(b)}{W(b)} K(b) \big[ c_1(b) J(b) + c_2(b) Y(b) \big],\\
c_2'(b) & = & +i \frac{J(b)}{W(b)} K(b) \big[ c_1(b) J(b)+ c_2(b) Y(b) \big].
\label{coupledsystem}
\end{eqnarray}
The large-$b$ boundary conditions are obtained for $A<0$ by choosing
$c_1=0$.  For $A>0$ we must examine which linear combination shrinks
under the influence of the $iK(b)$ term: we find $c_1=ic_2$ is the
shrinking solution.
At small $b$, $Y^2 \sim b^{2-2d}$, $W^{-1}\sim b^d$, and $K\sim b^2$, so
the equations are regular for $d<4$ and there is no obstacle to
solving them right down to $b=0$. The desired object, \Eq{wanttrans},
involves $\mathrm{Im}\,(c_1(0)/c_2(0))$.

For the NLO corrections one must perturb the original equations: $A\to A+\delta A$,
$K(b)\to K(b)+\delta K(b)$ and $\tilde f(b)\to \tilde f(b)+\delta \tilde f(b)$.
Then $\tilde f(b)$ still obeys \Eq{defgeneric} and acts as a source term
for $\delta f$:
\begin{equation}
\left(  \partial^2_b + (d/b) \partial_b + A + i K(b)  \right)\delta \tilde f(b) +\big[\delta A +\delta K(b)\big]\tilde f(b) = 0\,.
\label{defgenericNLO}
\end{equation}
Writing $\delta f = \delta c_1 J(b) + \delta C_2 Y(b)$ and applying an
analogous procedure, we find
\begin{eqnarray}
\nn \delta c_1'(b) & = & -\frac{Y(b)}{W(b)}
    \bigg\{ i K(b) \big[ \delta c_1(b) J(b) {+} \delta c_2(b) Y(b) \big]
 +(i\delta K(b){+}\delta A)\big[ c_1(b) J(b) {+} c_2(b) Y(b) \big]\bigg\},\\
\delta c_2'(b) & = & +\frac{J(b)}{W(b)} 
    \bigg\{iK(b) \big[ \delta c_1(b) J(b) {+} \delta c_2(b) Y(b) \big]
+(i\delta K(b){+}\delta A)\big[ c_1(b) J(b) {+} c_2(b) Y(b) \big]\bigg\}.\nn\\
&&\label{coupledsystemNLO}
\end{eqnarray}
When solving this numerically in the transverse case, one must take into account that
$\delta K(b\to0)\propto b$, which makes $\delta c'_1$ nonzero at the origin.

Note that both $\delta c_1(0)$ and $\delta c_2(0)$ give rise to a correction:
at linearized order
\begin{equation}
\mathrm{Im}\: \frac{c_1+\delta c_1}{c_2+\delta c_2}
 = \mathrm{Im}\: \frac{c_1}{c_2}
 + \mathrm{Im}\: \frac{c_2 \delta c_1 - c_1 \delta c_2}{c_2^2} \,.
\end{equation}

\section{Tabulated results}
\label{AppTab}

Here we present our strict leading-order and next-to-leading order
results, for $\nc=3$ and $\nf=3$ and several values of $k$.
Specifically, we present results for $\Ccoll$ and for $(1/g)\delta
\Ccoll$ for $k=3T,4T,5T,6T,8T,10T$ and a range of $k^-$ values, all
parametrized in terms of $k^-/g^2 T$. These results are independent
of the value of the coupling and can be readily plugged in 
Eqs.~\eqref{totallo} and \eqref{totnlo}.

\begin{table}
{\scriptsize \begin{center}
\begin{tabular}{|r|rrrrrr|} \hline
$k^-/g^2 T$ & $k=3T$ &  $k=4T$ &  $k=5T$ & 
 $k=6T$ &  $k=8T$ &  $k=10T$  \\ \hline
-3.00 & 10.6275 & 10.5329 & 10.5209 & 10.5360 & 10.5782 & 10.6133 \\ 
-2.80 & 9.8089 & 9.7295 & 9.7237 & 9.7413 & 9.7857 & 9.8217 \\ 
-2.60 & 8.9931 & 8.9292 & 8.9295 & 8.9498 & 8.9963 & 9.0335 \\ 
-2.40 & 8.1807 & 8.1323 & 8.1389 & 8.1620 & 8.2108 & 8.2492 \\ 
-2.20 & 7.3722 & 7.3395 & 7.3526 & 7.3786 & 7.4299 & 7.4697 \\ 
-2.00 & 6.5685 & 6.5517 & 6.5714 & 6.6004 & 6.6545 & 6.6959 \\ 
-1.80 & 5.7708 & 5.7701 & 5.7966 & 5.8288 & 5.8859 & 5.9292 \\ 
-1.60 & 4.9805 & 4.9964 & 5.0298 & 5.0655 & 5.1261 & 5.1715 \\ 
-1.40 & 4.2001 & 4.2328 & 4.2736 & 4.3130 & 4.3776 & 4.4256 \\ 
-1.20 & 3.4331 & 3.4832 & 3.5318 & 3.5753 & 3.6445 & 3.6958 \\ 
-1.00 & 2.6855 & 2.7536 & 2.8106 & 2.8588 & 2.9336 & 2.9891 \\ 
-0.90 & 2.3221 & 2.3995 & 2.4610 & 2.5117 & 2.5899 & 2.6479 \\ 
-0.80 & 1.9684 & 2.0553 & 2.1214 & 2.1750 & 2.2569 & 2.3179 \\ 
-0.70 & 1.6278 & 1.7242 & 1.7952 & 1.8519 & 1.9381 & 2.0025 \\ 
-0.60 & 1.3056 & 1.4116 & 1.4877 & 1.5476 & 1.6388 & 1.7073 \\ 
-0.50 & 1.0110 & 1.1261 & 1.2072 & 1.2706 & 1.3674 & 1.4408 \\ 
-0.40 & 0.7609 & 0.8831 & 0.9684 & 1.0352 & 1.1380 & 1.2173 \\ 
-0.35 & 0.6619 & 0.7857 & 0.8722 & 0.9403 & 1.0460 & 1.1286 \\ 
-0.30 & 0.5884 & 0.7111 & 0.7975 & 0.8662 & 0.9745 & 1.0605 \\ 
-0.25 & 0.5495 & 0.6661 & 0.7500 & 0.8182 & 0.9283 & 1.0179 \\ 
-0.20 & 0.5556 & 0.6578 & 0.7354 & 0.8011 & 0.9119 & 1.0055 \\ 
-0.15 & 0.6135 & 0.6886 & 0.7546 & 0.8157 & 0.9267 & 1.0260 \\ 
-0.10 & 0.7084 & 0.7448 & 0.7963 & 0.8532 & 0.9684 & 1.0786 \\ 
-0.05 & 0.7827 & 0.7899 & 0.8360 & 0.8968 & 1.0307 & 1.1653 \\ 
0.00 & 0.7902 & 0.8047 & 0.8663 & 0.9464 & 1.1246 & 1.3069 \\ 
0.05 & 0.7711 & 0.8170 & 0.9113 & 1.0266 & 1.2827 & 1.5507 \\ 
0.10 & 0.7618 & 0.8515 & 0.9947 & 1.1657 & 1.5512 & 1.9686 \\ 
0.15 & 0.7731 & 0.9217 & 1.1372 & 1.3937 & 1.9831 & 2.6365 \\ 
0.20 & 0.8106 & 1.0402 & 1.3590 & 1.7375 & 2.6070 & 3.5647 \\ 
0.25 & 0.8801 & 1.2181 & 1.6718 & 2.2012 & 3.3885 & 4.6623 \\ 
0.30 & 0.9872 & 1.4593 & 2.0663 & 2.7543 & 4.2529 & 5.8250 \\ 
0.35 & 1.1344 & 1.7540 & 2.5145 & 3.3536 & 5.1452 & 7.0049 \\ 
0.40 & 1.3177 & 2.0833 & 2.9880 & 3.9695 & 6.0458 & 8.1935 \\ 
0.50 & 1.7520 & 2.7830 & 3.9571 & 5.2162 & 7.8661 & 10.6018 \\ 
0.60 & 2.2199 & 3.4976 & 4.9391 & 6.4802 & 9.7162 & 13.0495 \\ 
0.70 & 2.6965 & 4.2218 & 5.9358 & 7.7638 & 11.5923 & 15.5264 \\ 
0.80 & 3.1799 & 4.9572 & 6.9473 & 9.0645 & 13.4877 & 18.0232 \\ 
0.90 & 3.6709 & 5.7033 & 7.9712 & 10.3785 & 15.3971 & 20.5340 \\ 
1.00 & 4.1694 & 6.4586 & 9.0051 & 11.7029 & 17.3169 & 23.0550 \\ 
1.20 & 5.1857 & 7.9906 & 11.0953 & 14.3742 & 21.1789 & 28.1185 \\ 
1.40 & 6.2221 & 9.5438 & 13.2067 & 17.0667 & 25.0613 & 33.2015 \\ 
1.60 & 7.2732 & 11.1120 & 15.3331 & 19.7738 & 28.9576 & 38.2980 \\ 
1.80 & 8.3355 & 12.6911 & 17.4702 & 22.4916 & 32.8642 & 43.4040 \\ 
2.00 & 9.4064 & 14.2788 & 19.6157 & 25.2174 & 36.7784 & 48.5175 \\ 
2.20 & 10.4840 & 15.8731 & 21.7676 & 27.9495 & 40.6986 & 53.6367 \\ 
2.40 & 11.5672 & 17.4727 & 23.9248 & 30.6868 & 44.6237 & 58.7606 \\ 
2.60 & 12.6549 & 19.0768 & 26.0863 & 33.4282 & 48.5529 & 63.8883 \\ 
2.80 & 13.7465 & 20.6846 & 28.2514 & 36.1732 & 52.4854 & 69.0192 \\ 
3.00 & 14.8412 & 22.2956 & 30.4196 & 38.9212 & 56.4207 & 74.1529 \\ 
\hline
\end{tabular} \end{center}}
\caption{Leading-order collinear contributions to the dilepton rate,
for $\nf=3$ QCD ($\nc=3$), for 6 values of momentum $k$ and several
virtualities. \label{LOtable}}
\end{table}

\begin{table}
{\scriptsize \begin{center}
\begin{tabular}{|r|rrrrrr|} \hline
$k^-/g^2 T$ & $k=3T$ &  $k=4T$   &  $k=5T$ &  $k=6T$ &
   $k=8T$ &  $k=10T$  \\ \hline
-3.00 & 1.2442 & 1.1773 & 1.1401 & 1.1158 & 1.0839 & 1.0621 \\ 
-2.80 & 1.2141 & 1.1482 & 1.1117 & 1.0878 & 1.0564 & 1.0347 \\ 
-2.60 & 1.1816 & 1.1170 & 1.0812 & 1.0577 & 1.0269 & 1.0055 \\ 
-2.40 & 1.1464 & 1.0833 & 1.0483 & 1.0254 & 0.9952 & 0.9741 \\ 
-2.20 & 1.1079 & 1.0465 & 1.0126 & 0.9904 & 0.9609 & 0.9403 \\ 
-2.00 & 1.0656 & 1.0063 & 0.9736 & 0.9522 & 0.9238 & 0.9037 \\ 
-1.80 & 1.0186 & 0.9619 & 0.9307 & 0.9103 & 0.8832 & 0.8638 \\ 
-1.60 & 0.9659 & 0.9124 & 0.8832 & 0.8640 & 0.8384 & 0.8200 \\ 
-1.40 & 0.9060 & 0.8566 & 0.8298 & 0.8123 & 0.7888 & 0.7717 \\ 
-1.20 & 0.8368 & 0.7928 & 0.7693 & 0.7541 & 0.7334 & 0.7181 \\ 
-1.00 & 0.7555 & 0.7190 & 0.7000 & 0.6878 & 0.6713 & 0.6585 \\ 
-0.90 & 0.7092 & 0.6775 & 0.6615 & 0.6514 & 0.6375 & 0.6264 \\ 
-0.80 & 0.6584 & 0.6326 & 0.6202 & 0.6126 & 0.6021 & 0.5932 \\ 
-0.70 & 0.6028 & 0.5844 & 0.5765 & 0.5720 & 0.5656 & 0.5594 \\ 
-0.60 & 0.5425 & 0.5333 & 0.5311 & 0.5305 & 0.5294 & 0.5268 \\ 
-0.50 & 0.4795 & 0.4819 & 0.4868 & 0.4912 & 0.4968 & 0.4986 \\ 
-0.40 & 0.4202 & 0.4368 & 0.4503 & 0.4607 & 0.4745 & 0.4820 \\ 
-0.35 & 0.3970 & 0.4211 & 0.4392 & 0.4528 & 0.4711 & 0.4817 \\ 
-0.30 & 0.3830 & 0.4140 & 0.4362 & 0.4529 & 0.4756 & 0.4895 \\ 
-0.25 & 0.3835 & 0.4190 & 0.4443 & 0.4634 & 0.4901 & 0.5073 \\ 
-0.20 & 0.4030 & 0.4383 & 0.4644 & 0.4849 & 0.5151 & 0.5357 \\ 
-0.15 & 0.4385 & 0.4678 & 0.4928 & 0.5143 & 0.5485 & 0.5739 \\ 
-0.10 & 0.4715 & 0.4947 & 0.5207 & 0.5455 & 0.5882 & 0.6221 \\ 
-0.05 & 0.4843 & 0.5134 & 0.5477 & 0.5810 & 0.6395 & 0.6873 \\ 
0.00 & 0.4969 & 0.5390 & 0.5862 & 0.6317 & 0.7119 & 0.7780 \\ 
0.05 & 0.5139 & 0.5741 & 0.6385 & 0.6995 & 0.8047 & 0.8890 \\ 
0.10 & 0.5346 & 0.6197 & 0.7045 & 0.7810 & 0.9030 & 0.9883 \\ 
0.15 & 0.5659 & 0.6778 & 0.7789 & 0.8613 & 0.9710 & 1.0264 \\ 
0.20 & 0.6098 & 0.7419 & 0.8444 & 0.9150 & 0.9886 & 1.0171 \\ 
0.25 & 0.6622 & 0.7973 & 0.8837 & 0.9357 & 0.9939 & 1.0367 \\ 
0.30 & 0.7130 & 0.8324 & 0.9013 & 0.9486 & 1.0260 & 1.0945 \\ 
0.35 & 0.7516 & 0.8508 & 0.9169 & 0.9742 & 1.0745 & 1.1538 \\ 
0.40 & 0.7750 & 0.8650 & 0.9402 & 1.0083 & 1.1195 & 1.2003 \\ 
0.50 & 0.8000 & 0.9019 & 0.9932 & 1.0700 & 1.1874 & 1.2730 \\ 
0.60 & 0.8243 & 0.9396 & 1.0374 & 1.1186 & 1.2452 & 1.3417 \\ 
0.70 & 0.8495 & 0.9732 & 1.0770 & 1.1640 & 1.3020 & 1.4091 \\ 
0.80 & 0.8741 & 1.0055 & 1.1158 & 1.2087 & 1.3570 & 1.4726 \\ 
0.90 & 0.8990 & 1.0377 & 1.1540 & 1.2520 & 1.4087 & 1.5308 \\ 
1.00 & 0.9243 & 1.0696 & 1.1910 & 1.2933 & 1.4567 & 1.5839 \\ 
1.20 & 0.9747 & 1.1305 & 1.2599 & 1.3685 & 1.5416 & 1.6762 \\ 
1.40 & 1.0231 & 1.1864 & 1.3214 & 1.4344 & 1.6142 & 1.7537 \\ 
1.60 & 1.0682 & 1.2371 & 1.3761 & 1.4923 & 1.6770 & 1.8202 \\ 
1.80 & 1.1099 & 1.2830 & 1.4251 & 1.5437 & 1.7321 & 1.8781 \\ 
2.00 & 1.1484 & 1.3248 & 1.4693 & 1.5898 & 1.7811 & 1.9293 \\ 
2.20 & 1.1840 & 1.3630 & 1.5094 & 1.6315 & 1.8251 & 1.9751 \\ 
2.40 & 1.2170 & 1.3982 & 1.5462 & 1.6695 & 1.8651 & 2.0165 \\ 
2.60 & 1.2477 & 1.4307 & 1.5801 & 1.7044 & 1.9016 & 2.0543 \\ 
2.80 & 1.2764 & 1.4609 & 1.6114 & 1.7367 & 1.9352 & 2.0890 \\ 
3.00 & 1.3034 & 1.4892 & 1.6406 & 1.7667 & 1.9664 & 2.1211 \\ 
\hline
\end{tabular} \end{center}}
\caption{Next-to-leading correction to the collinear contributions to
  the dilepton rate, for $\nf=3$ QCD ($\nc=3$), for 6 values of
  momentum $k$ and several virtualities. \label{NLOtable}}
\end{table}

We have also provided a $C$ code which evaluates these quantities for
input values of $k$ and $k^-$ (in units of $T$ and $g^2 T$), and
which allows the user to specify $\nf$ and $\nc$.  Users can use this
code to develop a table of results for interpolation.

\section{Extension to $k^-\sim gT$}
\label{app_gt2}

Here we will show that, if we apply our calculation outside of its range
of validity by considering $k^+ \simg T$ but $k^- \sim gT$, then our
leading-order calculation actually produces the correct leading and
next-to-leading contributions \textsl{if} we subtract off the free
collinear Born term and add back in the free, finite-angle Born term.

The leading-order contribution is given by evaluating the Born
diagram, \Fig{fig_born}.
Momentum conservation and the requirement that both propagators be on shell
forces $P$ to be semi-collinear, i.e. $P^2\sim gT^2$. Hence, thermal
masses can be dropped at LO
and the result is exactly the one given in \Eq{BornLOsemicoll}, which
is of order $\alpha K^2\sim \alpha gT^2$. We however note that, in obtaining
\Eq{BornLOsemicoll}, we have stretched $p^+$ to 0 and $-k^+$.
Since we encounter no singular behavior, we are allowed to do so at 
leading order, but it will require a subtraction in the soft region at
the next order to avoid a double-counting.

The next order, $\OO(\alpha g^2T^2)$, comes from the hard $\twotwo$,
semi-collinear and soft regions. As we showed at the end of
Sec.~\ref{sec:Born}, the purely collinear region $\pp\sim gT$, 
$p^+\sim T$  no longer contributes, as these processes are now
off shell, and collision corrections are suppressed by $\OO(
(\mm/-K^2)^2 )$. In the hard region, and thus for
$\pp^2\gg gT^2$, only  $\twotwo$ processes are kinematically
allowed. They are still not affected by the virtuality
in a first approximation, since $P^2\sim T^2\gg K^2$. However, one needs in 
principle to treat the approach to softer momenta with greater care, as
there are changes both to the kinematics and to the matrix elements. \Eq{softedge}
turns into
% Comment on this also regulating collinear divergence in the same way
\begin{eqnarray}
\Pi^<_{\twotwo,\sss{LO}}(K)
 \; \stackrel{\mathrm{small}\:p}{\longrightarrow}\;&&
\Bb \int_0 \pp d\pp \int_{-\infty}^{\infty} dp^+ 
  \frac{\theta(p^+-k^-/2+\pp^2/(2k^-))}{2(\pp^2+p^+{}^2)^{3/2}}\left(1+\frac{k^-p^+}{\pp^2}\right)\nn\\
   =&& \Bb \int_0 \frac{d\pp}{\pp},
   \label{softedge2}
\end{eqnarray}
where on the first line the $\theta$-function comes from accounting for $k^-$
in the kinematics in \Eq{Pi22LO}
and the extra term in round brackets likewise arises from a modification to
the Mandelstam variable $s$ for $p^+,\pp,k^-\sim gT$. Strikingly, the 
two corrections cancel, so that the approach to the soft sector is the
same as in \Eq{softedge}.  This cancellation relies on our choice to
leave the $\pp$ integration for last, which proves very convenient in
this case.  Cutting off the hard sector at a transverse regulator
$\mu_\perp$  gives
\begin{equation}
\label{twotwogT}
\Pi_{\twotwo,\sss{LO}}^<(K)  = \Bb \left[ \ln\left( \frac{T}{\mu_\perp}\right)
	  + C_\mathrm{hard}\left( \frac{k^+}{T} \right) \right].
\end{equation}

For what concerns the $\OO(g)$ corrections to the semi-collinear region, they come
from two sources. The first is by considering the next order in $k^-/k^+$. The replacement
of the collinear free term with the exact one, as introduced in \Eq{totallosubtr},
automatically takes into account all orders in that expansion. We can employ the same procedure
here. The second source comes 
about by considering the virtual cuts of the diagrams in \Fig{fig_diagrams}
for $P^2\sim gT^2$. These diagrams can be computed in a rather straightforward
way, since for a semicollinear $P$ the effect of the $Q$ integration is to give
rise to the on-shell limit of the hard self-energy and vertex correction,
which are equal to the same limit of their HTL counterparts, giving rise to terms
proportional to the asymptotic mass. In detail
\begin{eqnarray}
\Pi_{NLO,\mathrm{semi-coll}}^<(K) & = & \Bb
	\int_{-\infty}^\infty dp^+ 
        \frac{\nfd(k^+{+}p^+) [1-\nfd(p^+)]}{\nfd(k^+)}\frac{k^+}{2p^+(p^++k^+)}
	\nn \\
	 & & \hspace{-2.6cm}\times  \int_{\mu_\perp} \frac{d^2\pp}{(2\pi)^2}
	  \left[ \frac{k^+\pp^2}
            {2p^+(p^+{+}k^+)}\left(\frac{1}{p^+}-\frac{1}{p^++k^+}\right)
			 2\pi\delta'\left(\frac{k^+\pp^2}{2p^+(p^++k^+)}+k^-\right)\nn\right.\\
			\label{fpprobsc2}
&&\left.-\frac{2}{k^+} 2\pi \delta\left(\frac{k^+\pp^2}{2p^+(p^++k^+)}+k^-\right)\right]		
			  ,
\end{eqnarray}
where the $\delta'$ terms come from the self-energy diagrams and the $\delta$ term comes
from the vertex correction diagram. We are regulating the transverse integration
with $gT^2\gg \mu_\perp \gg g^2T^2$ to avoid overlap with the soft region. Using
\begin{eqnarray}
	&&\int_{\mu_\perp} \frac{d^2\pp}{2\pi} A \pp^2
            \delta'\left(A\pp^2+B\right)=
	\int_{\mu_\perp} \frac{d^2\pp}{2\pi}\pp^2\frac{d}{d\pp^2}\delta\left(A\pp^2+B\right)\nn\\
	&&\hspace{-1.2cm}=
	-\frac{\mu_\perp^2}{2}\delta(A\mu_\perp^2+B)-\int_{\mu_\perp^2}^\infty \frac{d\pp^2}{2}\delta(A\pp^2+B)
	=-\frac{\mu_\perp^2}{2}\delta(A\mu_\perp^2+B)-\frac{\theta(-B/A-\mu_\perp^2)}{2|A|},
	\label{deltamath}
\end{eqnarray}
where we have neglected the boundary term at $\pp=\infty$, as it cannot
contribute to the $p^+$ integration. We then obtain
\begin{eqnarray}
\Pi_{NLO,\mathrm{semi-coll}}^<(K) & = & \frac{\Bb}{2}
	\int_{-\infty}^\infty dp^+ 
        \frac{\nfd(k^+{+}p^+) [1-\nfd(p^+)]}{\nfd(k^+)}
	% \nn \\
% 	 & & \hspace{-2.6cm}\times
	  \left[ -\delta\left(\frac{p^+(p^++k^+)}{k^+}\right)
			 \nn\right.\\
			\label{fpprobscfinal}
&&\left.+\frac{(p^+)^2+(p^++k^+)^2}{p^+k^+(p^++k^+)}\theta\left(-\frac{2p^+(p^++k^+)}{k^+}k^--\mu_\perp^2\right)\right]		
			  .
\end{eqnarray}	
A comparison with \Eq{borntheta} shows that this amounts to 
having replaced $\mm$ with $\mu_\perp^2$ in the extrema of the integration. 
Hence, one can easily modify the results obtained for $\Pi^<_\mathrm{coll,Born,lim}$ 
and $\Pi^<_{\coll,\mathrm{Born},\theta}$  
in \Eq{borntotexpanded}, obtaining
\begin{equation}
	\label{finalnlosc}
	\Pi_{NLO,\mathrm{semi-coll}}^<(K)=\frac{\Bb}{2}\left[\ln\left(\frac{\mu_\perp^2}{k^+k^-}\right)-C_\mathrm{pair}\left(\frac{k^+}{T}\right)+\ln\frac{k^+}{2T}
	-1\right].
\end{equation}

\begin{figure}[ht]
  \begin{center}
   \begin{minipage}{0.13\textwidth}
   $\Pi^<_\mathrm{soft}(K)=$\\ \vspace{0.5cm}
   \end{minipage}
   \begin{minipage}{0.3\textwidth}
   \includegraphics[width=4cm]{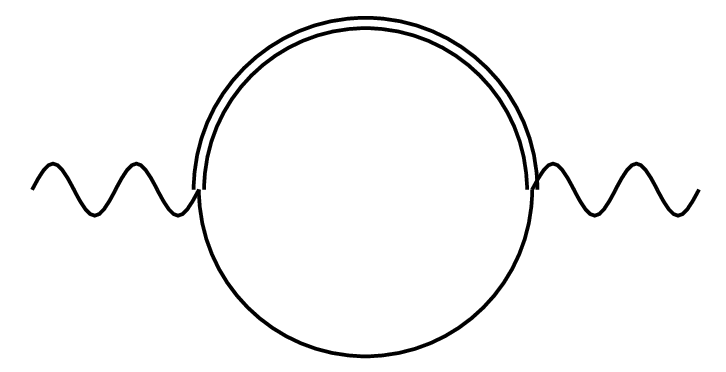}\\\vspace{0.1cm}
   \end{minipage}
   \includegraphics[width=\textwidth]{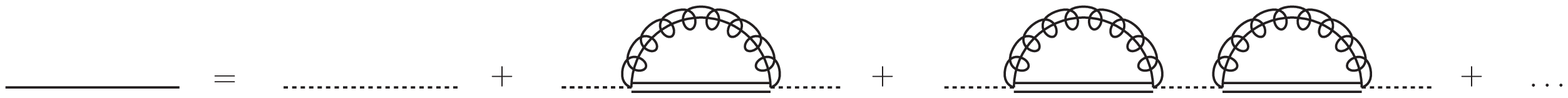}
  \end{center}
  \caption{The diagram contributing to the soft region at order $\alpha g^2T^2$. The double lines are the hard lines whereas 
  the dotted single line represent bare soft propagators. The single plain line is the HTL-resummed soft propagator. }
  \label{fig_lo_soft}
\end{figure}
For what concerns the soft sector, we have to evaluate the diagram in \Fig{fig_lo_soft},
where one line is hard, carrying the virtual photon's momentum, and the other is soft, $P\sim gT$.
The setup is very similar to the calculation for real photons and we refer to \cite{Ghiglieri:2013gia} 
for more details. In this case we find
\begin{eqnarray}
\label{leadingW}
 \Pi^{<}_{\mathrm{soft}}(K) = \frac{2\pi\Bb}{\mm}
  \int\frac{dp^+ d^2 \pp}{(2\pi)^3}
  \left[\left(1{-}\frac{p^z}{p}\right) \rho^+(P) 
         +\left(1{+}\frac{p^z}{p}\right)\rho^-(P) \right]_{p^-=-k^-}\! ,
\end{eqnarray}
where $\rho^{\pm}(P)=S_R^\pm(P)-S_A^\pm(P)$ are the spectral function
of the $+$ and $-$ modes of the fermion HTL propagator. The $1\mp p^z/p$ arise from the
Dirac trace and $p^-=-k^-$ stems from the requirement that $P+K$ be on shell. This 
is the only difference with the $k^-\siml g^2T$ case, which includes the real photon, 
where $p^-\sim g^2T$ can be taken to be zero, as we have remarked in Sec.~\ref{sec:NLOvirt}.
As in that case, however, we are free to exploit the fact that retarded (advanced)
functions are analytic in the upper (lower) half-plane for any timelike and lightlike
variable. Thus, we are free to deform the integration contour for $p^+$ away from the 
real axis on arcs at large $|p^+|$, as depicted in Fig.~13 of \cite{Ghiglieri:2013gia}.
Along these arcs the retarded and advanced functions in \Eq{leadingW} expand to
\begin{eqnarray}
&&\left(1-\frac{p^z}{p}\right)S^+(p^+,-k^-,\pp)+\left(1+\frac{p^z}{p}\right)S^-(p^+,-k^-,\pp)
\bigg\vert_{\vert p^+\vert\to+\infty}\nn\\
  &=&\frac{i}{p^+}\left(-\frac{\pp^2}{2p^+k^-+\pp^2+\mm}+1\right)=\frac{i}{p^+}+\order{\frac{1}{(p^+)^2}}.
\label{upperarc}
\end{eqnarray}
Since we have regulated the higher-lying regions with a IR cutoff on $\pp$ only,
this means that in the soft region we should use the same $\mu_\perp$ as an UV cutoff
and stretch the $p^+$ integration to infinity, which effectively renders only the
$1/p^+$ term relevant. We then see a major difference with the $k^-\siml g^2T$
case, where also the first term in round brackets on the second line would have contributed.
In the present case, on the other hand, the resulting $i/p^+$, once integrated over
$dp^+$, gives rise to a $\mu_\perp^2$ behavior, i.e.
\begin{equation}
\label{leadingWfinal}
 \Pi^{<}_{\mathrm{soft}}(K) = \frac{2\pi\Bb}{\mm}
  \int^{\mu_\perp}\frac{ d^2 \pp}{(2\pi)^2}=\frac{\Bb}{2\mm}\mu_\perp^2.
\end{equation}
However, as we mentioned before, when deriving the leading-order
semi-collinear result, we have integrated for $p^+$ and $p_\perp$
down to $0$, using bare propagators. Hence we should subtract the bare limit
here to avoid double countings. It is easy to see (just take $\mm$ to 0 in
\Eq{upperarc}) that the bare and resummed results coincide in this case, and
hence the contribution from the soft sector vanishes after the subtraction.

This apparently puzzling result can be understood as a cancellation
between the spacelike and timelike regions, which correspond to the 
soft limits of the real and virtual corrections respectively to the Born term.
Previous calculations in the literature \cite{Thoma:1997dk,Carrington:2007gt}
have dealt with these two regions separately and with different regularization schemes,
which might make the cancellation incomplete.
Hence, the dependence on $\mu_\perp$ has to cancel between the two IR
divergences in the hard $\twotwo$ and semi-collinear regions. 
Indeed, the $\OO(g)$ correction to \Eq{BornLOsemicoll} is obtained by
summing \Eq{twotwogT} and \eqref{finalnlosc}, yielding 
 \begin{equation}
 	\label{finalnlogt}
 	\delta\Pi_{NLO,k^-\sim gT}^<(K)=\frac{\Bb}{2}\left[\ln\left(\frac{T^2}{k^+k^-}\right)+2C_\mathrm{hard}\left(\frac{k^+}{T}\right)-C_\mathrm{pair}\left(\frac{k^+}{T}\right)+\ln\frac{k^+}{2T}
 	-1\right],
 \end{equation}
 so that the complete result for $k^-\sim gT$, up to order $\alpha g^2T^2$, reads
\begin{equation}
	\Pi_{NLO,k^-\sim gT}^<(K)=\Pi^<_\mathrm{free}(K)+\delta\Pi_{NLO,k^-\sim gT}^<(K),
\end{equation} 
which includes higher orders in $k^-/k^+\sim g$ in $\Pi^<_\mathrm{free}$. These can be subtracted if needed.
Hence, as we set out to prove, this corresponds to the expansion for large
$k^-$ of $\Pi^<_{LO,\mathrm{large}}(K)$.


\begin{thebibliography}{39}

\bibitem{Ghiglieri:2013gia} 
  J.~Ghiglieri, J.~Hong, A.~Kurkela, E.~Lu, G.~D.~Moore and D.~Teaney,
  %``Next-to-leading order thermal photon production in a weakly coupled quark-gluon plasma,''
  JHEP {\bf 1305}, 010 (2013)
  [arXiv:1302.5970 [hep-ph]].
  %%CITATION = ARXIV:1302.5970;%%

\bibitem{highEphotons}
S.~Afanasiev {\it et al.}  [PHENIX Collaboration],
  %``Measurement of Direct Photons in Au+Au Collisions at $\sqrt{s_{NN}} = 200$ GeV,''
  Phys.\ Rev.\ Lett.\  {\bf 109}, 152302 (2012)
  [arXiv:1205.5759 [nucl-ex]];
  %%CITATION = ARXIV:1205.5759;%%
  %37 citations counted in INSPIRE as of 26 Sep 2014
A.~Adare {\it et al.} (PHENIX collaboration),
  %``Direct photon production in $d+$Au collisions at $\sqrt{s_{NN}}=200$ GeV,''
  Phys.\ Rev.\ C {\bf 87}, 054907 (2013)
  [arXiv:1208.1234 [nucl-ex]].
  %%CITATION = ARXIV:1208.1234;%%
  %17 citations counted in INSPIRE as of 26 Sep 2014
 M.~Wilde [ALICE Collaboration],
  %``Measurement of Direct Photons in pp and Pb-Pb Collisions with ALICE,''
  Nucl.\ Phys.\ A {\bf 904-905}, 573c (2013)
  [arXiv:1210.5958 [hep-ex]].
  %%CITATION = ARXIV:1210.5958;%%
  %44 citations counted in INSPIRE as of 26 Sep 2014

\bibitem{EXPdilepton}
A.~Adare {\it et al.}  [PHENIX Collaboration],
  %``Enhanced production of direct photons in Au+Au collisions at $\sqrt{s_{NN}}=200$ GeV and implications for the initial temperature,''
  Phys.\ Rev.\ Lett.\  {\bf 104}, 132301 (2010)
  [arXiv:0804.4168 [nucl-ex]];
  %%CITATION = ARXIV:0804.4168;%%
  %286 citations counted in INSPIRE as of 26 Sep 2014
A.~Adare {\it et al.}  [PHENIX Collaboration],
  %``Detailed measurement of the $e^+ e^-$ pair continuum in $p+p$ and Au+Au collisions at $\sqrt{s_{NN}} = 200$ GeV and implications for direct photon production,''
  Phys.\ Rev.\ C {\bf 81}, 034911 (2010)
  [arXiv:0912.0244 [nucl-ex]].
  %%CITATION = ARXIV:0912.0244;%%
  %205 citations counted in INSPIRE as of 26 Sep 2014

\bibitem{McLerran:1984ay} 
  L.~D.~McLerran and T.~Toimela,
  %``Photon and Dilepton Emission from the Quark - Gluon Plasma: Some General Considerations,''
  Phys.\ Rev.\ D {\bf 31}, 545 (1985).
  %%CITATION = PHRVA,D31,545;%%  

\bibitem{CKMSY}
S.~Caron-Huot, P.~Kovtun, G.~D.~Moore, A.~Starinets and L.~G.~Yaffe,
  %``Photon and dilepton production in supersymmetric Yang-Mills plasma,''
  JHEP {\bf 0612}, 015 (2006)
  [hep-th/0607237].
  %%CITATION = HEP-TH/0607237;%%
  %131 citations counted in INSPIRE as of 04 Aug 2014

\bibitem{AGMZ}
  P.~Aurenche, F.~Gelis, G.~D.~Moore and H.~Zaraket,
  %``Landau-Pomeranchuk-Migdal resummation for dilepton production,''
  JHEP {\bf 0212}, 006 (2002)
  [hep-ph/0211036].
  %%CITATION = HEP-PH/0211036;%%

\bibitem{Laine:2013vma} 
  M.~Laine,
  %``NLO thermal dilepton rate at non-zero momentum,''
  JHEP {\bf 1311}, 120 (2013)
  [arXiv:1310.0164 [hep-ph]].
  %%CITATION = ARXIV:1310.0164;%%	

\bibitem{Ghisoiu:2014mha} 
  I.~Ghisoiu and M.~Laine,
  %``Interpolation of hard and soft dilepton rates,''
  JHEP {\bf 1410}, 83 (2014)
  [arXiv:1407.7955 [hep-ph]].
  %%CITATION = ARXIV:1407.7955;%%

\bibitem{latt_people}
O.~Kaczmarek, E.~Laermann, M.~M\"uller, F.~Karsch, H.~T.~Ding, S.~Mukherjee, A.~Francis and W.~Soeldner,
  %``Thermal dilepton rates from quenched lattice QCD,''
  PoS ConfinementX {\bf }, 185 (2012)
  [arXiv:1301.7436 [hep-lat]];
  %%CITATION = ARXIV:1301.7436;%%
  
\bibitem{Ansatzmethod}
 O.~Kaczmarek and M.~M\"uller,
  %``Temperature dependence of electrical conductivity and dilepton rates from hot quenched lattice QCD,''
  PoS LATTICE {\bf 2013}, 175 (2013)
  [arXiv:1312.5609 [hep-lat]].
  %%CITATION = ARXIV:1312.5609;%%
  %4 citations counted in INSPIRE as of 26 Sep 2014

\bibitem{MEM}
 F.~Karsch, S.~Datta, E.~Laermann, P.~Petreczky, S.~Stickan and I.~Wetzorke,
  %``Hadron correlators, spectral functions and thermal dilepton rates from lattice QCD,''
  Nucl.\ Phys.\ A {\bf 715}, 701 (2003)
  [hep-ph/0209028].
  %%CITATION = HEP-PH/0209028;%%
  %55 citations counted in INSPIRE as of 26 Sep 2014

\bibitem{Arnold:2001ms} 
  P.~B.~Arnold, G.~D.~Moore and L.~G.~Yaffe,
  %``Photon emission from quark gluon plasma: Complete leading order results,''
  JHEP {\bf 0112}, 009 (2001)
  [hep-ph/0111107].
  %%CITATION = HEP-PH/0111107;%%  

\bibitem{Arnold:2001ba} 
  P.~B.~Arnold, G.~D.~Moore and L.~G.~Yaffe,
  %``Photon emission from ultrarelativistic plasmas,''
  JHEP {\bf 0111}, 057 (2001)
  [hep-ph/0109064].
  %%CITATION = HEP-PH/0109064;%%

\bibitem{GaleMajumder}
A.~Majumder and C.~Gale,
  %``Dileptons from a quark gluon plasma with finite baryon density,''
  Phys.\ Rev.\ D {\bf 63}, 114008 (2001)
  [Erratum-ibid.\ D {\bf 64}, 119901 (2001)]
  [hep-ph/0011397].
  %%CITATION = HEP-PH/0011397;%%
  %20 citations counted in INSPIRE as of 13 Aug 2014

\bibitem{Gervais:2012wd} 
  H.~Gervais and S.~Jeon,
  %``Photon Production from a Quark-Gluon-Plasma at Finite Baryon Chemical Potential,''
  Phys.\ Rev.\ C {\bf 86}, 034904 (2012)
  [arXiv:1206.6086 [nucl-th]].
  %%CITATION = ARXIV:1206.6086;%%
  
\bibitem{Kapusta:1991qp} 
  J.~I.~Kapusta, P.~Lichard and D.~Seibert,
  %``High-energy photons from quark - gluon plasma versus hot hadronic gas,''
  Phys.\ Rev.\ D {\bf 44}, 2774 (1991)
  [Erratum-ibid.\ D {\bf 47}, 4171 (1993)].
  
\bibitem{Baier:1991em} 
  R.~Baier, H.~Nakkagawa, A.~Niegawa and K.~Redlich,
  %``Production rate of hard thermal photons and screening of quark mass singularity,''
  Z.\ Phys.\ C {\bf 53}, 433 (1992).
  %%CITATION = ZEPYA,C53,433;%%    

\bibitem{Braaten}
E.~Braaten and R.~D.~Pisarski,
  %``Soft Amplitudes in Hot Gauge Theories: A General Analysis,''
  Nucl.\ Phys.\ B {\bf 337}, 569 (1990).
  %%CITATION = NUPHA,B337,569;%%
  %1061 citations counted in INSPIRE as of 30 Sep 2014

\bibitem{Frenkel}
J.~Frenkel and J.~C.~Taylor,
  %``High Temperature Limit of Thermal QCD,''
  Nucl.\ Phys.\ B {\bf 334}, 199 (1990).
  %%CITATION = NUPHA,B334,199;%%
  %455 citations counted in INSPIRE as of 30 Sep 2014

\bibitem{Aurenche:1998nw} 
  P.~Aurenche, F.~Gelis, R.~Kobes and H.~Zaraket,
  %``Bremsstrahlung and photon production in thermal QCD,''
  Phys.\ Rev.\ D {\bf 58}, 085003 (1998)
  [hep-ph/9804224].
  %%CITATION = HEP-PH/9804224;%%  

\bibitem{Zakharov}
B.~G.~Zakharov,
  %``Fully quantum treatment of the Landau-Pomeranchuk-Migdal effect in QED and QCD,''
  JETP Lett.\  {\bf 63}, 952 (1996)
  [hep-ph/9607440];
  %%CITATION = HEP-PH/9607440;%%
  %364 citations counted in INSPIRE as of 13 Aug 2014
%``Light cone path integral approach to the Landau-Pomeranchuk-Migdal effect,''
  Phys.\ Atom.\ Nucl.\  {\bf 61}, 838 (1998)
  [Yad.\ Fiz.\  {\bf 61}, 924 (1998)]
  [hep-ph/9807540].
  %%CITATION = HEP-PH/9807540;%%
  %142 citations counted in INSPIRE as of 13 Aug 2014

\bibitem{Aurenche:2002pd} 
  P.~Aurenche, F.~Gelis and H.~Zaraket,
  %``A Simple sum rule for the thermal gluon spectral function and applications,''
  JHEP {\bf 0205}, 043 (2002)
  [hep-ph/0204146].
  %%CITATION = HEP-PH/0204146;%%
	
\bibitem{Carrington:2007gt} 
  M.~E.~Carrington, A.~Gynther and P.~Aurenche,
  %``Energetic di-leptons from the Quark Gluon Plasma,''
  Phys.\ Rev.\ D {\bf 77}, 045035 (2008)
  [arXiv:0711.3943 [hep-ph]].
  %%CITATION = ARXIV:0711.3943;%%	

\bibitem{CaronHuot:2008ni} 
  S.~Caron-Huot,
  %``O(g) plasma effects in jet quenching,''
  Phys.\ Rev.\ D {\bf 79}, 065039 (2009)
  [arXiv:0811.1603 [hep-ph]].
  %%CITATION = ARXIV:0811.1603;%%

\bibitem{CaronHuot:2008uw} 
  S.~Caron-Huot,
  %``On supersymmetry at finite temperature,''
  Phys.\ Rev.\ D {\bf 79}, 125002 (2009)
  [arXiv:0808.0155 [hep-th]].
  %%CITATION = ARXIV:0808.0155;%%      

\bibitem{Aurenche:2002pc} 
  P.~Aurenche, F.~Gelis and H.~Zaraket,
  %``Enhanced thermal production of hard dileptons by 3 ---> 2 processes,''
  JHEP {\bf 0207}, 063 (2002)
  [hep-ph/0204145].
  %%CITATION = HEP-PH/0204145;%%  

\bibitem{CaronHuot:2009ns} 
  S.~Caron-Huot,
  %``Asymptotics of thermal spectral functions,''
  Phys.\ Rev.\ D {\bf 79}, 125009 (2009)
  [arXiv:0903.3958 [hep-ph]].
  %%CITATION = ARXIV:0903.3958;%%

\bibitem{Thoma:1997dk} 
  M.~H.~Thoma and C.~T.~Traxler,
  %``Production of energetic dileptons with small invariant masses from the quark - gluon plasma,''
  Phys.\ Rev.\ D {\bf 56}, 198 (1997)
  [hep-ph/9701354].
  %%CITATION = HEP-PH/9701354;%%

\end{thebibliography}
\end{document}